\documentclass[a4paper,fleqn,usenatbib]{mnras1}

\usepackage[T1]{fontenc}
\usepackage{ae,aecompl}

\usepackage{float}

\usepackage{graphicx}
\usepackage{mathtools}
\usepackage[mathscr]{euscript}

\usepackage{bm}
\usepackage{deluxetable}

\usepackage{ulem}
\title[Short title, max. 45 characters]{MNRAS \LaTeXe\ template -- title goes here}

\title[Cosmic rays and galaxy cluster dynamics]{Manufacturing cosmic rays in the evolving dynamical states of galaxy clusters}
\author[R S John, S Paul et al.]{Reju Sam John$^{1,2,3}$, Surajit Paul$^{2,3,5}$
\thanks{E-mail: surajit@physics.unipune.ac.in}, Luigi Iapichino$^{4}$, Karl Mannheim$^{5}$
\newauthor{and Harish Kumar$^{1}$}\\
$^{1}$Department of Physics, Pondicherry Engineering College, Puducherry, 605014, India\\
$^{2}$Department of Physics, SP Pune University, Pune, 411007, India \\
$^{3}$Inter-University Centre for Astronomy and Astrophysics, Pune, 411007, India\\
$^{4}$Leibniz-Rechenzentrum der Bayerischen Akademie der Wissenschaften, Boltzmannstr. 1, D-85748 Garching b. Munchen, Germany\\
$^{5}$Lehrstuhl f\"ur Astronomie, Institut f\"ur Theoretische Physik und Astrophysik,
Universit\"at W\"urzburg, Emil Fischer-Str. 31,\\ D-97074 W\"urzburg, Germany}

\begin{document}

\date{Accepted . Received ; in original form }

\pagerange{\pageref{firstpage}--\pageref{lastpage}} \pubyear{2002}

\maketitle

\label{firstpage}

\begin{abstract}

Galaxy clusters are known to be reservoirs of Cosmic Rays (CRs), as inferred from theoretical calculations or detection of CR-derived observables. CR acceleration in clusters is mostly attributed to the dynamical activity that produces shocks. Shocks in clusters emerge out of merger or accretion, but which one is more effective in producing CRs? at which dynamical phase? and why? To this aim, we study the production or injection of CRs through shocks and its evolution in the galaxy clusters using cosmological simulations with the {\sc enzo} code. Particle acceleration model considered here is primarily the Diffusive Shock Acceleration (DSA) of thermal particles, but we also report a tentative study with pre-existing CRs. Defining appropriate dynamical states using the concept of virialization, we studied a sample of merging and non-merging clusters. We report that the merger shocks (with Mach number $\mathcal{M}\sim2-5$) are the most effective CR producers, while high-Mach peripheral shocks (i.e. $\mathcal{M}>5$) are mainly responsible for the brightest phase of CR injection in clusters. Clusters once merged, permanently deviate from CR and X-ray mass scaling of non-merging systems, enabling us to use it as a tool to determine the state of merger. Through a temporal and spatial evolution study, we found a strong correlation between cluster merger dynamics and CR injection. We observed that the brightest phase of X-ray and CR injection from clusters occur respectively at about 1.0 and 1.5 Gyr after every mergers, and CR injection peaks near to the cluster virial radius (i.e $r_{200}$). Delayed CR injection peaks found in this study deserve further investigation for possible impact on the evolution of CR-derived observables from galaxy clusters.

\end{abstract}


\section{Introduction}\label{intro}

Being on the top of the large-scale structure mass hierarchy, galaxy clusters, during its emergence and Giga years (Gyr) of evolution time, witness a cascade of dynamical events. Studies have confirmed that the galaxy clusters had emerged out of the aggregation of structures like galaxies, groups of galaxies, hot inter-cluster medium and warm hot materials from the filaments \citep{Bykov2015SSRv,Yu2015ApJ,Springel2006Natur}. The energy budget of galaxy clusters is thus the cumulative effect of all possible dynamical events such as star formation, supernova activity, AGN activity, galaxy formation and structure mergers at all levels (galaxy, groups, clusters). Recently, a possible role of Dark Matter (DM) annihilation has been claimed \citep{2013A&A...550A.134P,Huber2013A&A,Berezinsky_1997ApJ}. So, galaxy clusters are among the best available cosmic laboratories \citep{Blasi_2004JKAS,Berezinsky_1997ApJ,Huber2013A&A} to test the energy evolution of the universe. Though a major fraction of the energy in the clusters is thermal in nature, a non-negligible non-thermal component is also present \citep{Brunetti_2014IJMPD,Petrosian2008SSRv}. 
The observations of cluster-scale radio emission provide the most important evidence for non-thermal components in galaxy clusters \citep{Brunetti_2014IJMPD,Weeren_2019SSRv}.

Galaxy clusters usually grow by continuous accretion and mergers of smaller and bigger groups of galaxies. Mergers of galaxy clusters in the mass range between $10^{14} M_{\odot}$ and $10^{15} M_{\odot} $ release an enormous binding energy of about $10^{63}$ to  $10^{65}$ ergs. It has been proposed that the energy released during these mergers creates huge pressure gradient in the cluster core (see e.g.  \cite{PAUL2012IoP}). As a result, the medium starts expanding supersonically inducing strong shocks in the baryonic gas \citep{1996ApJ...456..422K,2000PhDT.........7M,Paul2011ApJ,Kang2013ApJ}. Merger energy then gets dissipated mainly through shock heating and turbulence stirring in the medium \citep{Sarazin_1987_ApJ,Dolag2005MNRAS,Iapichino_2010AIP,Bourdin2013ApJ}.  Moreover, due to shock acceleration, a significant amount of charged particles gains energy and reaches the non-thermal regime by converting the shock energy and thus contributing to the CR population \citep{Berrington2003ApJ,Sarazin2001cghr,Jones2001AIPC}. As energy gets dissipated, clusters tend to attain a state of virialisation. At the point of virialisation, almost an equipartition of gravitational energy to kinetic energy can be observed.  Roughly a thermodynamic equilibrium is also expected to be established in the cluster medium, if a sufficiently long time is elapsed before it reaches the virialisation \citep{Ballesteros_2006MNRAS}. Clusters deviating from this equilibrium are either having more potential energy (usually this is the phase of a rapid mass accretion/merger) or having more  kinetic energy in the course of relaxation (i.e. a state of disturbance or higher entropy) \citep{Barsanti_2016A&A}. 

Usually shocks due to mergers emerge at the central core of the newly formed clusters and travels radially as spheroidal wave-front towards the virial radius and beyond \citep{Paul2011ApJ,Iapichino_2017MNRAS,van_Weeren_2011MNRAS,PAUL2012IoP,Bagchi_2006Sci,Bagchi_2011ApJ}. Time-scale of such dissipation for a single merger is found to be about 1-2 Gyr \citep{Paul2011ApJ,Roettiger_ApJ_1999} indicating an energy dissipation rate of the order of 10$^{\rm{47}}$   erg s$^{-1}$. In this process, the thermal particles in the medium are injected at the shocks and a fraction of them gets accelerated to high energy CR particles through collision-less shock energy dissipation \citep{Malkov1998PhRvE,Ellison1985ApJ}. Also, pre-existing high-energy particles originated from other processes such as numerous star-formation events, supernovae explosions, AGN activity etc., can further be accelerated to very high energies through multiple shock crossings \citep{Kang2007APh}. So, it is quite evident that a fraction of these particles from the thermal pool gets converted into non-thermal CRs, mainly via diffusive shock acceleration (DSA; \;\citealt{1987PhR...154....1B}, \citealt{2001RPPh...64..429M}). The cosmic ray production due to shock acceleration is then expected to be more prominent during the post-merger violent relaxation phase. This indicates a possible strong connection of CR injection and the dynamical state of the cluster systems \citep{Brunetti_2014IJMPD}. 

Compared to CR electrons, CR protons have larger radiative cooling time, quantitatively by the square of the mass ratio, $(m_p/m_e)^2$ and are estimated to be of the order of 3-30 Gyr. Having such a high cooling time, protons accelerated by accretion shocks at the outskirts of the cluster can accumulate in the galaxy clusters spanning the Hubble time \citep{HONG2014ApJ,Volk1996SSRv}. 
So, clusters are expected to be major source of high-energy CRs. But surprisingly, no firm detection of gamma rays from clusters are reported yet. However, the non detection of gamma-ray emission from galaxy clusters (for the most recent data, see \citealt{Ackermann_2016ApJ}) puts constraints on the content of CR protons in the cluster centres (\citealt{Brunetti_2017MNRAS} and references therein). 

In this paper, we study the production and evolution of CR particles in the merging and non-merging galaxy clusters to understand the brightest phase of CR injection in clusters. Specifically, we try to understand what are the main factors that control the CR injection in clusters, in terms of the link between dynamical states of clusters to their energy distribution, and to observable diagnostics like the X-ray luminosity. We use the concept of virialisation (see Section~\ref{dyna-stat}) and merger energy distribution (see Section~\ref{virial-evol}) to characterise the dynamical state of the simulated galaxy clusters. The simulations, run with the {\sc{Enzo}} code \citep{Bryan_2014ApJS} includes the effects of star formation, supernova (SN) feedback and cooling due to radiative processes. For our cosmological simulations of structure formation, AGN feedback has not been considered as it is unlikely to have significant impact on scales of high-mass objects like galaxy clusters \citep{LeBrun_2014MNRAS,McCarthy_2010MNRAS}. Also, the only mechanism considered for CR acceleration in this work is the first-order Fermi process, primarily for thermally accelerated particles. Though keeping in mind the possible effect of pre-existing CRs (preCR hereafter), we also present a special case of acceleration of preCRs as a tentative study. In all our calculations, we neglect the second-order Fermi acceleration and explicit contribution from SNs and AGNs.

Our paper is organised as follows. After giving the introduction in Section~\ref{intro}, we will start with the description of our simulations in Section~\ref{sim-dtls}.  Modelling thermal and non-thermal energies is described in Section~\ref{energy-compute} and sample selection will be introduced in Section~\ref{sample}. Evolution of dynamical parameters is explained in Section~\ref{dynamical-states-ev}. In Section~\ref{res-dic}, we report the findings and discuss the results.  Linking energetics of clusters with merging events and to the dynamical states will also be reported in this section. Finally, we will summarise the outcomes in Section~\ref{sum}. 

\section[Simulations]{Simulations of cluster mergers and its analysis details}\label{sim-dtls}

To create a sample of galaxy clusters, simulations were performed with the Adaptive Mesh Refinement (AMR), grid-based hybrid (N-body plus hydrodynamical) code {\sc Enzo} v.~2.2 \citep{O'Shea_2005_Springer,Bryan_2014ApJS}. This code uses adaptive refinement in space and time, and introduces non-adaptive refinement in mass by multiple child grid insertions. A flat $\Lambda$-CDM background cosmology  with specific cosmology parameters $\Omega_\Lambda$ = 0.7257, $\Omega_m$ = 0.2743, $\Omega_b$ = 0.0458,  h = 0.702 and primordial power spectrum normalisation $\sigma_8$ = 0.812 has been used. Cosmology parameters are obtained from $\Lambda$-CDM cosmology, derived from WMAP (5-years data) combined with the distance measurements from the Type Ia supernovae (SN) and the Baryon Acoustic Oscillations (BAO) (see \cite{Komatsu_2009_ApJS}). Simulations have been initialised at redshift $z = 60$ using the \cite{Eisenstein1999ApJ} transfer function, and evolved up to $z = 0$. An ideal equation of state was used for the gas, with $\gamma = 5/3$. 

Our research mainly focuses on the production of CRs through Diffusive Shock Acceleration (DSA), which is a strong function of shock strength \citep{1987PhR...154....1B,Ryu_2003ApJ}. Detection and quantification of shocks is thus an important part of this project. Merger shocks compress the medium near to the shock front, but post shock medium suffers rapid expansion and can induce adiabatic as well as radiative cooling \citep{Akahori_2012_PASJ,Medvedev_2009ApJ,Choi_2004_ApJ}. We have thus used the cooling and heating model of \citet{Sarazin_1987_ApJ} that takes into account the effect of radiative cooling due to X-ray emissions and heating due to star formation, star motions and SN. Futher, we have used the star formation and feedback schemes of \citet{Cen1992ApJL} with a feedback of 0.25 solar. We have named this model with additional physics as `coolSF' runs (for details, see \citet{Paul_2017MNRAS}). Shocks have been detected in our simulations using un-split velocity jump method of \citet{Vazza_2011MNRAS,Vazza_2009MNRAS,Skillman2008ApJ} with a temperature floor of $10^4$K which is found to give better results in AMR simulations \citep{Vazza_2011MNRAS}.

Since shocks are a vital component of our study, in our adaptive mesh refinement strategy, we have used the refinement criteria based on shocks along with the over-density. Over-density criteria has been used on both the DM and the baryon component. The cells will be flagged for refinement if the local density $\rho_\text{i}$, where `i' can indicate either baryons or DM, fulfils the following criterion: 
\begin{equation}
\label{threshold}
\rho_\text{i} > f_\text{i} \rho_0 \Omega_\text{i} N^l
\end{equation}
where $\rho_0 = 3 H_0^2 / 8 \pi G$ is the critical density, $H_0$ is the Hubble parameter at the present epoch, $\Omega_\text{i}$ are the cosmological density parameters for either baryons or DM, and the refinement factor is $N=2$. Here $l$ is the refinement level (for root grid, $l =0$). In the current work we set  $f_\text{i} = 4$ for both the overdensities $f_{\text{b}} = f_{\text{DM}} = 4$. Since density is very low on the outskirts of the large-scale objects compared to their innermost parts, the AMR based on overdensity is not suitable for refining the flow there \citep{Iapichino_2017MNRAS}. For this reason we also require an additional AMR criterion, whose effectiveness does not depend strongly on density. We use therefore the refinement on shocks as additional AMR criterion. The Mach number of a shock is not a direct function of density, but it depends on the ratio of post and pre-shock densities. This is the  reason why, even at regions far away from the centre, where density is very low, we could still reach up to the highest level of refinement at the shocked regions. Enstrophy, derived from vorticity field has been earlier used by \citet{Iapichino_2017MNRAS} for similar purposes.

We have first produced several low-resolution, DM-only runs to select suitable clusters depending on their mass, size and dynamical history and re-simulated them at high resolutions with the full physics setup. The evolution and mass accretion of the forming structures have been followed by producing merger trees with the yt toolkit \citep{Turk_2011ApJS}. Our main simulations are of (128 Mpc h$^{-1})^3$ volume and utilise a root grid with $64^3$ elements. We have introduced 2 nested child grids. Furthermore, 4 additional AMR levels are used in the central (32 $\rm{Mpc})^3$ volume. The effective spatial resolution is thus of about 30 kpc for the simulations of our reference set (`RefRES' hereafter). We have performed a resolution study using some lower and higher resolution and different root grid resolution simulations compared to our RefRES. For a detailed resolution study please refer Appendix~\ref{appen-res-study}.

\section{Modelling X-ray and cosmic ray emissions}\label{energy-compute}
\subsection{Thermal X-ray emissions model}

Merger-induced shocks, gas drag by the density clumps and other similar mechanisms can  heat up the intra-cluster medium (ICM) up to few times $10^8$K \citep{Sarazin_2002ASSL,Sarazin_1986_RvMP}. The ICM consequently emits X-rays through thermal bremsstrahlung (for a review; \citealt{Felten_1966ApJ,Sarazin_1986_RvMP}), Inverse Compton Scattering of Cosmic Microwave Background Radiation photons by relativistic electrons present in the ICM \citep{Costain_1972ApJ,Brecher_1972ApJ} and also by thermal synchrotron radiation. In our simulations, X-ray luminosity and emissivity for a given photon energy range are computed from the emission tables created by  ~\cite{Smith_2008MNRAS} with the photoionization code {\sc cloudy}~\citep{Ferland_1998PASP}, that includes all the above mentioned emission processes. We have computed the emission fields for photon energy range from $0.1 \; \rm{keV}$ to $12.0 \;\rm{keV}$, which is the typical range for most X-ray telescopes. This range also includes more than 99\% of bolometric luminosity. From our simulations, using the above mentioned relations, we have computed the temperature-weighted X-ray emissions. 
 
\subsection{Non-thermal cosmic ray emission model}\label{non-therm}

It is well known that collision-less shocks accelerate CRs mainly via the DSA process as studied extensively for supernovae remnant and inter-planetary shocks  \citep{Krymskii_1977DoSSR,Drury_1983RPPh,1987PhR...154....1B,Blasi_2013A&ARv}. CR acceleration in galaxy clusters has been so far studied through simulations by solving the diffusion-convection equation explicitly for gas dynamical shocks \citep{Pfrommer2008MNRAS,Kang2013ApJ,Kang2007ApJ}, as post-processing models implemented in cosmological simulations \citep{HONG2014ApJ} and with dynamical feedback and time evolution computations of CRs in cosmological setup \citep{Vazza_2016MNRAS,Miniati_2001ApJ}. All these simulations as well as upper limits from observations \citep{Ackermann_2016ApJ,Ackermann_2014ApJ,Huber2013A&A} show that the CR pressure in ICM can go upto only a few percent of the gas thermal pressure of the systems.

In this work, we model the acceleration of CR protons by injection of available energised particles at the cluster shocks as a post process of cosmological hydrodynamic simulation. Cluster shocks are the collisionless shocks, mediating interactions of charged particles and magnetic fields in the tenuous and high $\beta = P_{th}/P_{B} (\sim 100)$ cosmic plasmas \citep{Kang2007ApJ}, where $P_{th}$ and $P_{B}$ are the thermal and Magnetic pressure in the ICM. Injected particle population in our scheme contains particles from the thermal pool. It may as well contain the preCR particles due to historical energetic events in the system as discussed in the Section~\ref{intro}, which we have taken as a separate case at the end of this study (See Section~\ref{pre-exist}). These energised charged particles get accelerated by the collision less shocks, mainly via Fermi~I process or DSA. In this process, due to multiple crossings of charged particles across the shocks, the shock flow energy gets transferred to the shocked particles \citep{Drury_1983RPPh}. A fraction of the shock kinetic energy flux is thus assumed to be transferred to the CR energy flux. The kinetic energy to CR energy flux conversion efficiency of the shocks is defined as $\eta(\mathcal{M})$ which is a function of the Mach number of the shock $\mathcal{M}$. The CR energy flux at the shock surface can therefore be expressed as
\begin{equation}
f_{CR} = \eta (\mathcal{M}) \times f_{kin}
\end{equation}

where $\mathcal{M}$ is the shock Mach number,  $f_{kin}$ is the kinetic energy flux and given by  $f_{kin} = \frac{1}{2}\rho v^3$ for a density $\rho$ and velocity of the medium. Expressing Mach number of the shock as $\mathcal{M}~=~v/c_{s}$ with flow velocity $v$ and sound speed in the medium as $c_s$, kinetic energy flux can be written as a function of shock Mach number $\mathcal{M}$. The CR energy flux then becomes a steep function of Mach number only and can be expressed as 
\begin{equation}\label{eq:fcr}
f_{CR}= \eta(\mathcal{M}) \times \frac{1}{2}\rho(\mathcal{M} c_{s})^3
\end{equation}

\begin{figure}
\includegraphics[width=0.5\textwidth]{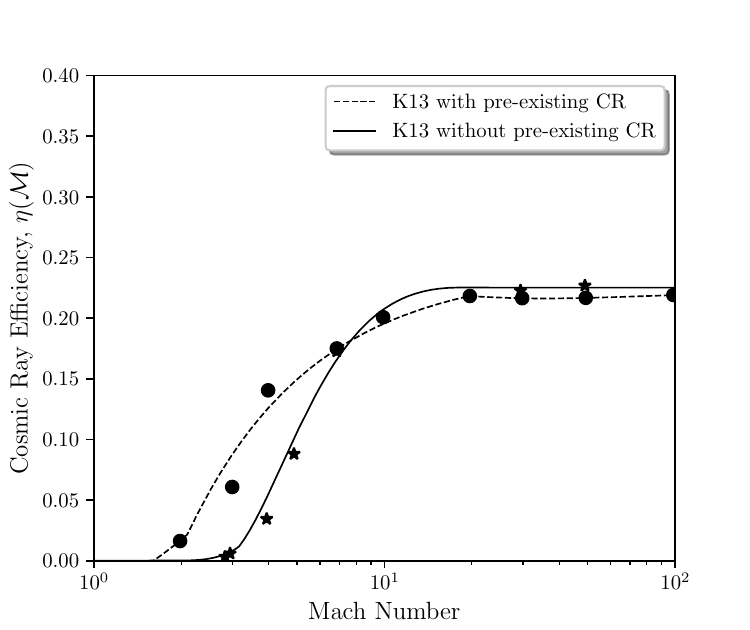}
\caption{CR acceleration efficiency $\eta$ as a function of Mach number ($\mathcal{M}$) in the model of \citet{Kang2013ApJ} has been plotted for without preCRs (solid curve) as well as with preCR population (dashed curve).} \label{fig:mach-effi}
\end{figure}

\begin{figure}
\includegraphics[width=0.49\textwidth]{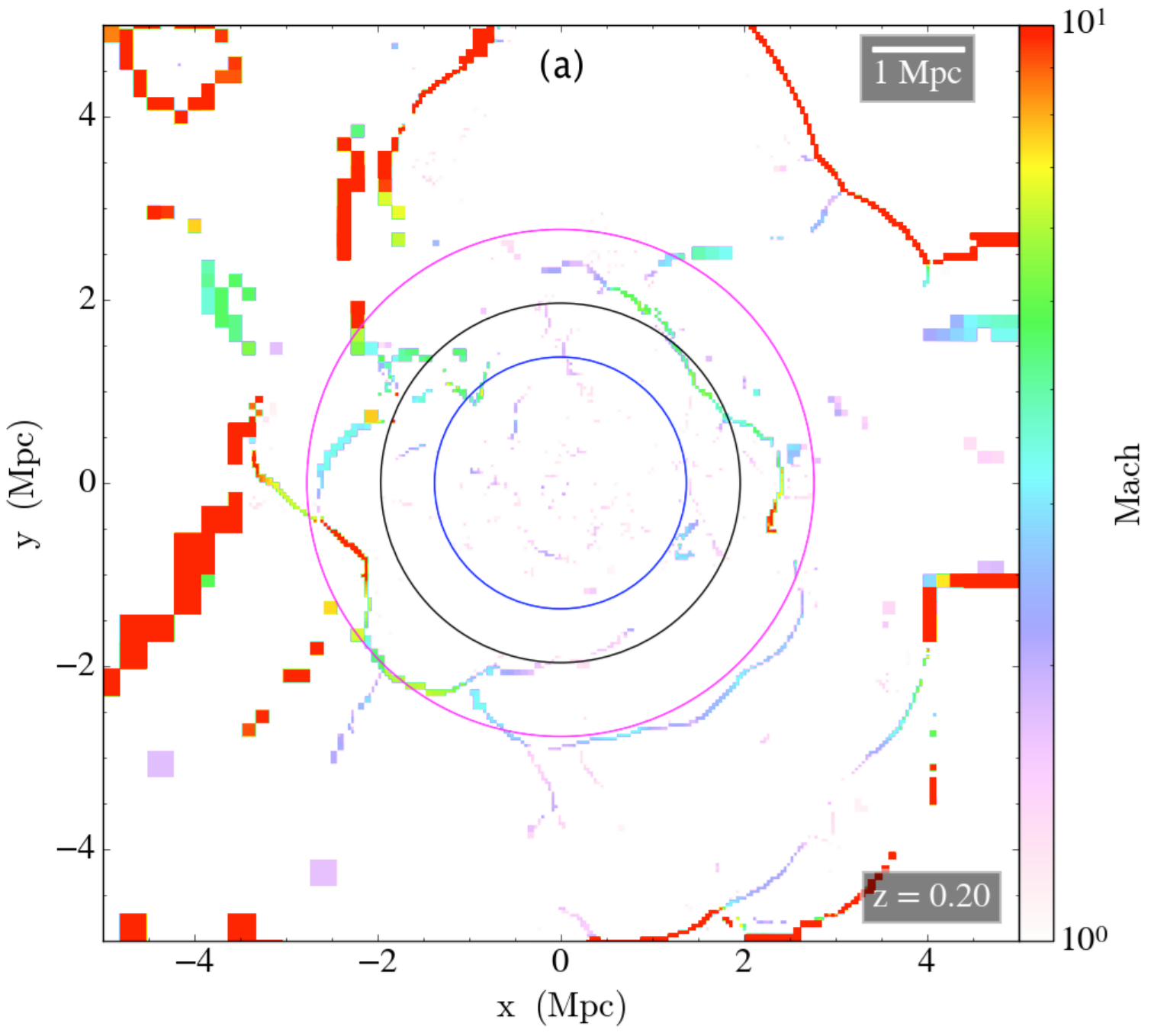}\\
\includegraphics[width=0.49\textwidth]{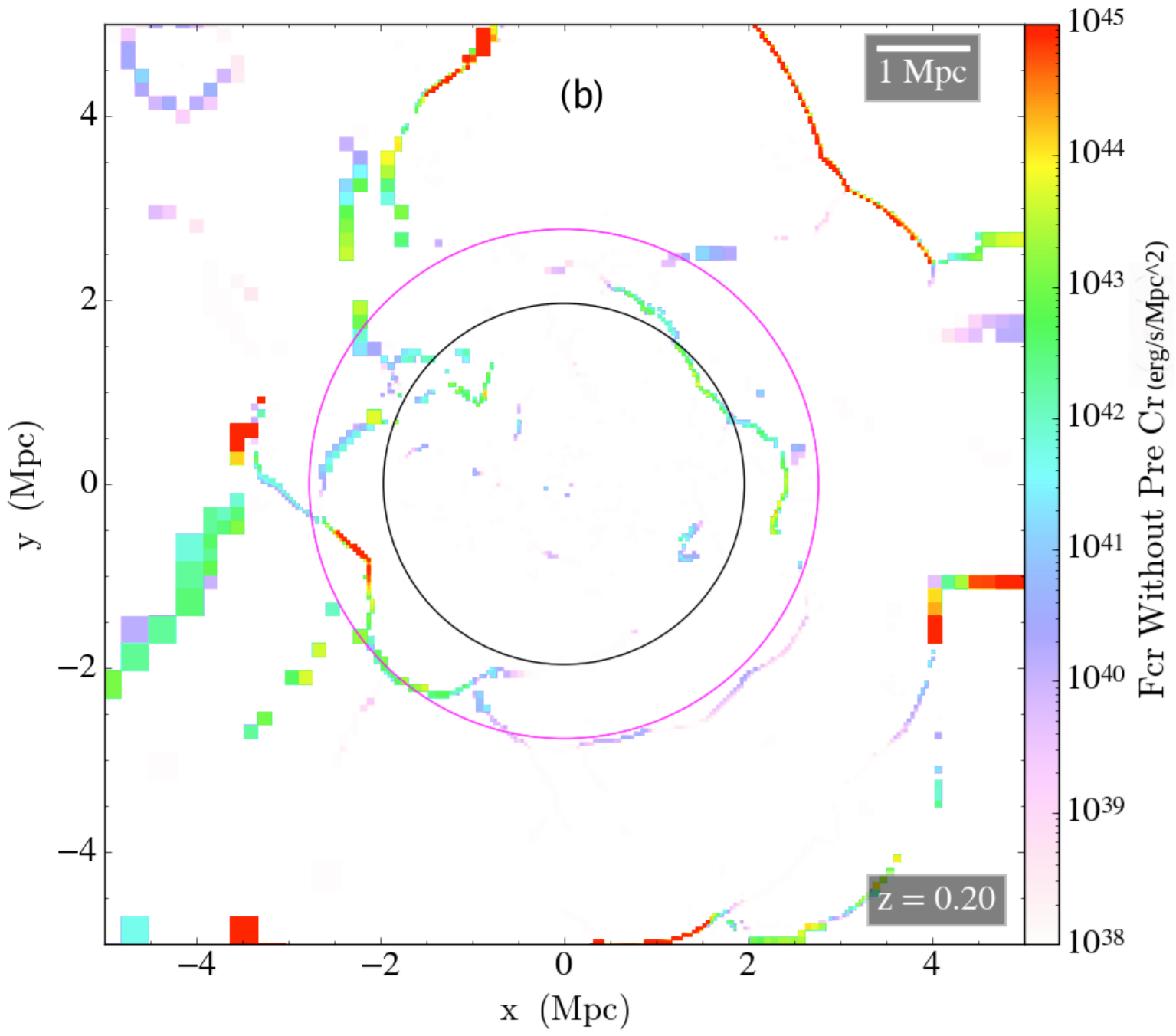}
\caption{{\bf Panel (a):} Mach number of the shocks is plotted in a slice of size of 10 Mpc $h^{-1}$ each side. Three concentric circles represent the radius at overdensity 1000 ($r_{1000}$, blue) i.e. core, 500 ($r_{500}$, black) and 200 ($r_{200}$, magenta) i.e. virial radius. {\bf Panel (b):} Same as panel (a), but showing CR flux ($F_{CR}$ in $erg\;s^{-1}\;\rm{Mpc}^{-2}$). Here, two circles indicate $r_{500}$ (black) and $r_{200}$ (Magenta).} \label{fig:mach-CRF}
\end{figure}

Injection of ambient energetic particles, acceleration at collision-less shocks and their transport in the ICM are not very well understood processes. So, deriving an analytical form of $\eta(\mathcal{M})$ is rather a complicated task. It has been recognised that a part of the supra-thermal particles available in the Maxwellian distribution tail successfully swim against the Alfv\'en waves further advecting downstream, and leak to the upstream across the shock and get injected as the CR population \citep{Ryu_2003ApJ,Kang2007APh,Kang2013ApJ}. This model is known as the ``thermal leakage'' injection model. Further, magnetic field amplification happens due to CR streaming instabilities in the shock precursor and the Alfv\`enic drift of scattering centers. This play a significant role in DSA at astrophysical shocks such as supernova remnant shocks (e.g., \citealt{Lucek_2000MNRAS,Bell_2004MNRAS,Bykov_2014ApJ}). Self-amplification of magnetic fields and fast Alfv\`enic drift in the shock precursor has been implemented into the standard DSA model by \citet{Kang2013ApJ}, who explored various parameters responsible for CR acceleration by solving 1-D diffusion-convection equation to generate the CR acceleration efficiency as a function of Mach number (i.e. $\eta(\mathcal{M})$). It is found that the inclusion of magnetic field amplification in the acceleration model leads to a lower efficiency (saturated $\eta(\mathcal{M})=0.225$ than the earlier reported value in \citet{Kang2007ApJ} (saturated $\eta(\mathcal{M})=0.55$). Another model with hybrid-simulations by \citealt{Caprioli_2014ApJ} shows further lower (about $10^{-1}$) saturated efficiency  for above Mach number $\mathcal{M}>5$. However they do not discuss the details of acceleration efficiency at low Mach numbers as well as re-acceleration scenario of preCRs. 

Our study focuses on Mach-dependent CR acceleration and evolution in galaxy clusters using the models of \citet{Kang2013ApJ}. Piece-wise analytical functions of efficiencies with Mach numbers have been derived from \citet{Kang2013ApJ}. A comparison plot of efficiencies of conversion of kinetic to CR flux at shocks according to this model have been provided for both acceleration of thermal particles as well as the re-acceleration of preCRs in Figure~\ref{fig:mach-effi}. It can be noticed in this plot, also, pointed out by \citet{Kang2013ApJ} that preCR particles contribute significantly more at low Mach numbers ($\mathcal{M} < 5$). However, the acceleration efficiency converges to the same value at $\mathcal{M}>7$. Therefore, the role of preCRs would not be significant for high-Mach ($\mathcal{M}>5$) shocks. Low-Mach shocks are mostly found well inside the clusters. Instead,  cluster outskirts are dominated by shocks beyond Mach number $\mathcal{M} >5$, because of mergers of substructures as well as infall of gas along the filaments \citep{Skillman2008ApJ,Vazza_2009MNRAS,PAUL2012IoP}. This is visually shown in Figure~\ref{fig:mach-CRF}(a) (see also \citealt{Ryu_2003ApJ,Miniati_2001ApJ}). These are the shocks that are also known for producing the highest energy CRs \citep{HONG2014ApJ}.

To study the possible connection of dynamical states with CR injection, an important factor is to understand the brightest phase of CR production in clusters. Our study therefore mainly focuses on the particles accelerated from the thermal pool by the strong shocks. Further, to asses the importance of preCR in this context, in the absence of time evolution process in our simulations, only a special case of re-acceleration of preCRs is discussed. We would also like to stress that in our model, we consider the acceleration as an instantaneous process. The acceleration is treated as a model ingredient but still in post-processing, because we do not implement any physical model to properly follow the time evolution of CRs in our simulations.

\begin{figure}
\includegraphics[width=0.48\textwidth]{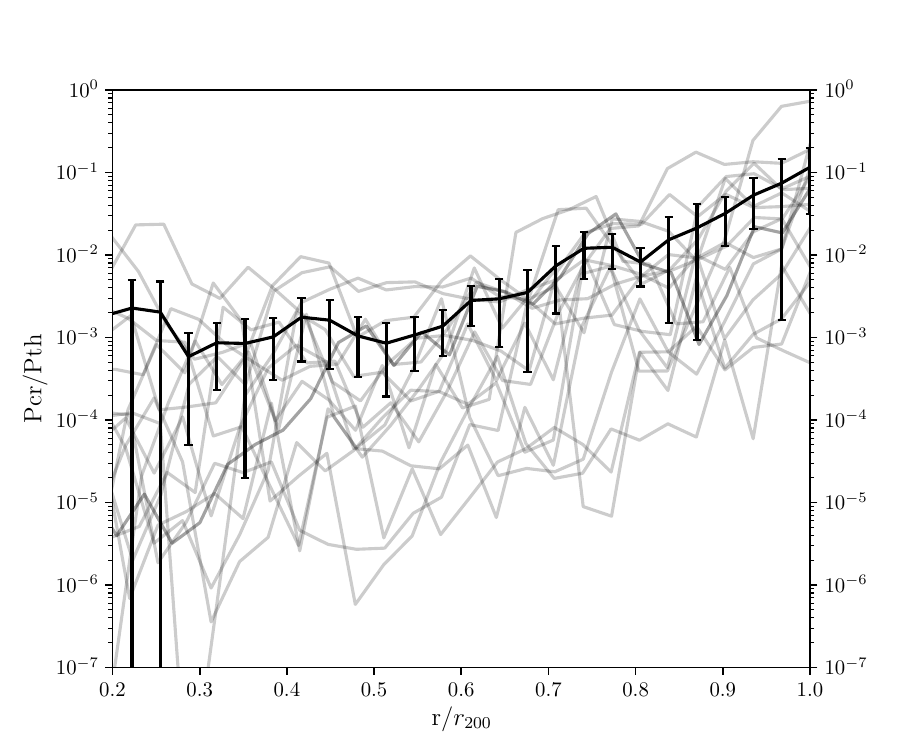}
\caption{Radial variation of ratio of CR to thermal pressure ($X(r)=P_{CR} (r \le r_{200})/P_{th} (r\le r_{200})$) has been plotted for high mass ($>5\times 10^{14}M_{\odot}$) merging clusters showing their redshifts evolution after mergers.}\label{fig:pcr_by_pth}
\end{figure}

The energy of the CRs injected into the cluster medium after shock acceleration has been computed as the fraction of shock kinetic energy flux
\begin{equation}
E_{CR_{inj}}= \eta(\mathcal{M})\times\frac{1}{2}\rho(\mathcal{M}.c_s)^2\times(\Delta x)^3
\end{equation}
whereas thermal energy of the cell is defined as 
\begin{equation}
E_{th}= \frac{3}{2} Nk_BT\times(\Delta x)^3
\end{equation}
where $\Delta x$ is the width of each computed cell, $N$ and $T$ are respectively the number of particles and temperature of the cell. CR injection to gas thermal pressure is further computed by taking ratio of average CR pressure and gas pressure within increasing radii i.e. 
\begin{equation}
X(r)=P_{CR} (r \le r_{200})/P_{th} (r \le r_{200})
\end{equation}
where the pressure of CRs is $P_{CR} = (\Gamma_{CR} - 1) E_{CR}$ with $\Gamma_{CR}$ as 4/3 for a ultra-relativistic equation of state for CRs. Similarly, $P_{th} = (\Gamma_{th} - 1) E_{th}$ with equation of state of an ideal gas $\Gamma_{th}$ as 5/3. The ratio $P_{CR}/P_{th}$ has been plotted for the simulated high mass (final mass $> 5\times 10^{14} M_{\odot}$) merging clusters during their evolution after the first merger in Figure~\ref{fig:pcr_by_pth}. It can be noticed in our plot that the ratio goes to as high as 70\% at maximum and beyond radius $r_{500}$ at almost near to $r_{200}$. The average profile shows that till $0.7r_{200}$, the CR injection pressure is much below 1\% of the thermal pressure, but beyond $0.75r_{200}$ it increases fast and it reaches almost 11\% at the outskirts (near to $r_{200}$) of the clusters. Although we are aware that our assumption of an instantaneous CR acceleration in post-processing might be flawed when the CR energy content gets large, compared to the thermal one, here we see that in most of the inner regions of the clusters, this is not the case.

\subsubsection{Procedure of CR emission computation in our simulations}\label{cr-compt-model}

Shocked cells in our simulations have been flagged using the un-split velocity jump method described in  \citet{Vazza_2011MNRAS,Vazza_2009MNRAS,Skillman2008ApJ}.
The CR energy flux in the cell is computed from Equation~\ref{eq:fcr} using the cell values for the required variables. For typical clusters, the roughly estimated value for cosmic ray flux ($F_{CR}$) from a shocked cell will be almost $10^{40-42}$ erg s$^{-1} \rm{Mpc}^{-2}$ (see Fig.~\ref{fig:mach-CRF}(b)). This corresponds to a CR luminosity of about $10^{43}$ erg s$^{-1}$ from the total surface area of a typical merger shock in our simulation, corroborating the estimation by \citet{2003PhPl...10.1992S}. This calculation is valid for an ideal case where a single shock has emerged from a major merger of two groups. In the more realistic scenario probed by the numerical simulations, most of the mergers either pass through the phase of core oscillations producing multiple shocks or multiple mergers take place within a short period of time producing numerous shock surfaces inside the clusters. Moreover, we consider only the shock surfaces that contains shocks with Mach number $\mathcal{M}>1.1$. To capture all these shocks and their contribution to the CR production, we have considered all those shock surfaces and computed the cell volume weighted average CR flux over the whole volume of the cluster. Finally, we have computed the total CR luminosity for the whole virial volume of the clusters using the relation

\begin{equation}
 L_{CR} =  4 \;\pi \; r_{vir}^2 \; f_{CR}
\end{equation}

Where $r_{vir}$ is the virial radius at over density 200 of each clusters. This will capture all the shocks within the virial ($r_{200}$) volume while computing the total luminosity. To nullify the effect of our complicated distribution of AMR cell sizes, luminosities within the said regions are computed as weighted average quantities using 

\begin{equation}
L_{CR_{wt}}=\frac{\sum_i V_i\; F_{CR}}{\sum_i V_i}
\end{equation}

where $V_i$ is the volume of the $i^{th}$ cell.

In rare cases ($<1\%$ cases), filamentary inroads are observed within $r_{200}$ (e.g. Fig.~\ref{fig:mach-CRF}) of the clusters as the virial radius is computed assuming a spherical symmetry. These filamentary entries being placed near to almost voids, bring previously unprocessed materials inside these virial spheres. In absence of photoionization process in our simulation, this may produce extremely high Mach numbers ($>>10$) in few cells. Such cells observed to produce unphysical CR flux ($F_{CR}$) (see Fig.~\ref{fig:mach-CRF}, $F_{CR}>10^{45}\; \rm{erg\;s^{-1}\;Mpc^{-2}}$). Preliminary numerical tests have suggested us to suppress the CR acceleration above a Mach number threshold of M=50. Therefore, the shocked cells with Mach number above that value are removed from our calculations.

\section{Sample selection}\label{sample}

\begin{table*}
\caption{List of highly refined clusters in our simulation samples. Total mass and the baryonic mass of the listed clusters are given in the $2^{nd}$ and $3^{rd}$ column respectively. These are the masses within the radius $r_{200}$ and radius $r_{500}$, which is given in the $4^{th}$ column. Similarly, temperatures are given in the $5^{th}$ column. Finally, the merging state of each of these clusters is reported in the $6^{th}$ column.}\label{table_cluster}
\begin{tabular}{cccccl}
\hline
Run & Virial Mass (T) & Virial Mass (B) & Virial Radius & Virial Temperature & Merging State\\ 
 & ($10^{14}\rm{ M_{\sun}}$) & ($10^{14} \rm{M_{\sun}}$) & (Mpc) & ($10^{7}$K) & \\
 & $r_{200} / r_{500}$ & $r_{200} / r_{500}$ & $r_{200} / r_{500}$ & $r_{200} / r_{500}$ & \\
 \hline
$Cl_1$   &  24.81/21.03 & 3.62/3.05  & 3.70/2.76 & 4.73/6.21 & Non-merging \\
$Cl_2$   & 15.67/12.58  & 2.23/1.76  & 3.17/2.27  &  4.13/5.05 & Merging   \\
$Cl_3$   & 9.08/7.78  & 1.34/1.12  & 2.64/1.96  & 2.87/3.74 & Merging \\
$Cl_4$   &  9.41 /7.57  & 1.29/1.05  & 2.74/1.97   & 2.58/3.06 & Merging \\
$Cl_5$   &  9.41/7.33  &  1.31/1.04 & 2.70/1.97 & 2.70/3.41 & Non-merging\\
$Cl_6$   & 7.26/6.67 &0.96/0.86 & 2.57/1.83 & 2.46/3.91 & Merging\\
$Cl_7$   & 6.99/5.77 & 0.97/0.81 & 2.42/1.80& 2.24/2.82 & Merging \\
$Cl_8$   & 6.72/5.55&0.94/0.75 & 2.47/1.74 & 1.84/2.73 & Non-merging\\
$Cl_9$   & 0.89/0.78 &0.11/0.09 & 1.21/0.91 & 0.59/0.73 & Merging\\
$Cl_{10}$   & 0.88/0.72 & 0.12/0.10& 1.23/0.88 & 0.58/0.82 & Non-merging \\
\hline
\end{tabular}
\end{table*}

For this study, we have simulated ten galaxy clusters with `CoolSF' physics (Section~\ref{sim-dtls}). From five different realisations of the low-resolution, DM-only setup, we have chosen the objects that in our judgement seemed to be better representative of different mass ranges and dynamical states. The cluster mass in our samples ranges from a lowest of $\rm{10^{13}\; M_{\odot}}$ to the highest of $\rm{>\;10^{15}\; M_{\odot}}$, providing almost two orders of magnitude in mass to study a wide range of systems (for details of these simulated clusters see Table~\ref{table_cluster}). 

For identifying objects in our simulations, we have used virial radius  as a delimiter, where `virial' refers to the quantity at over density of 200 to the critical density of the universe at that redshift. In the following, over density should be understood as compared to the critical density only. Though we have focused on a central cluster in each high-resolution run, a large number of smaller objects are available within the (32 $\rm{Mpc)^3}$ central volume of our simulations  where AMR has been allowed. So, along with the main 10 clusters, for statistical studies, we have a few hundred clusters with the same resolution by combining the data from different sets of simulations. The number has further been increased by considering the output snapshots from various redshifts. We made sure that the DM particles and baryon gas that forms the objects of interest are mostly coming from the well refined Lagrangian region. To do so, we have taken the objects that are placed only within the central (20 $\rm{Mpc)^3}$ volume. Though in our simulations, we have galaxy groups with mass about 10$^{13}\;\rm{M_{\odot}}$, it has been observed that low mass systems does not follow the cluster mass scaling and possibly needs physical modelling of smaller scales \citep{Bharadwaj_2015A&A,Surajit2015,Paul_2017MNRAS}. So, to ensure roughly the cluster properties of these systems as well as enough statistics and mass range, the minimum mass we have considered for our study is $5\times 10^{13}\; \rm{M_{\odot}}$. We did not put any cut-off on highest mass and our list contains mass upto $2\times 10^{15}\; \rm{M_{\odot}}$. Mass resolution at the innermost child grid in our simulations is less than $10^{9}\; \rm{M_{\odot}}$ i.e. with more than $10^4$ particles provide enough resolution for even the smallest objects used. Also, with 30 kpc effective resolution, systems above $5\times 10^{13} \; \rm{M_{\odot}}$ having $r_{200}$ above 500 kpc, get adequately resolved in space.

\subsection{Sampling by dynamical activities}\label{dyna-act-samp}

\subsubsection{Merging and non-merging systems}\label{mer-non-mer}

In our high resolution simulation sample of 10 clusters ($Cl_1$ to $Cl_{10}$ in Table~\ref{table_cluster}), we have chosen 6 merging systems and 4 non-merging systems. We have computed the merger trees for these objects and accordingly indicated the merging and non-merging systems. A non-merging, relaxed galaxy cluster is a system that has not experienced any merger with a mass ratio larger than 0.1 almost for the last  8 Gyr (from redshift $z=1.0$ to the current epoch ($z=0$). i.e. almost all its evolution time (e.g. see the column~6 in Table~\ref{table_cluster}). All other systems are labelled as mergers. Merging systems are spanning from small mergers (ratio more than 0.1 but less than 0.3)  to major mergers (ratio more than 0.3, for definition see: \citet{Paul2011ApJ} and references therein). Also, there is a cluster that got merged only once (e.g. $Cl_7$ ) and a system with multiple mergers (e.g. $Cl_2$). We have chosen the biggest two i.e. $Cl_1$ (non-merging) and $Cl_2$ (merging) as the primary representative of non-merging and merging samples as their final mass is almost identical (about $2\times 10^{15} \; \rm{M_{\odot}}$). The masses of all other clusters in the list are chosen very carefully to compare them well. Each of the non-merging systems has at least one comparable merging system with similar mass. These clusters are mainly used to study the time evolutions of different physical parameters. We have used different pairs of systems for comparing different parameters depending upon requirements (e.g. mass range, number of mergers etc.). 

\subsubsection{Dynamical states of galaxy clusters using the concept of virialisation}\label{dyna-stat}

Evolving galaxy clusters are dynamically very active and cannot attain virial equilibrium in a time shorter than their dynamical time-scale \citep{Ballesteros_2006MNRAS}. For a self gravitating system like the clusters, the corrected form of virial theorem can be expressed as below \citep{Davis_2011MNRAS}.

\begin{equation}
\frac{1}{2}\frac{d^2 I}{dt^2}\; =\; 2 K + W - E_s
\end{equation}

Where $I$ is the moment of inertia of the system and $K$ and $W$ represent the kinetic energy and gravitational potential energy of the system respectively. For the systems like the galaxy clusters, where influence of external potential (i.e. outside the virialised radius) is significant, $W$ can be expressed as $U_{int} + U_{ext}$, where $U_{int}$ is the internal potential of the studied system and $U_{ext}$ is due to the mass outside the cluster radius (here, $r_{200}$), but whose tidal effect can be felt. The surface pressure term $E_s$ comes from the velocity dispersion of the galaxies that creates an extra outwards force on the cluster virial surface.  At the time when the second derivative of $I$ vanishes, the system said to have attained the virial equilibrium \citep{Ballesteros_2006MNRAS,Davis_2011MNRAS}.

To study the virialisation of galaxy clusters, we define virial ratio as 

 \begin{equation}
\mathcal{ R}_1 = \frac{(U_{int} + U_{ext}-E_s)}{2\times{KE} }
\label{vir-res1}
 \end{equation}
 
 \begin{figure}
\includegraphics[width=1.1\columnwidth]{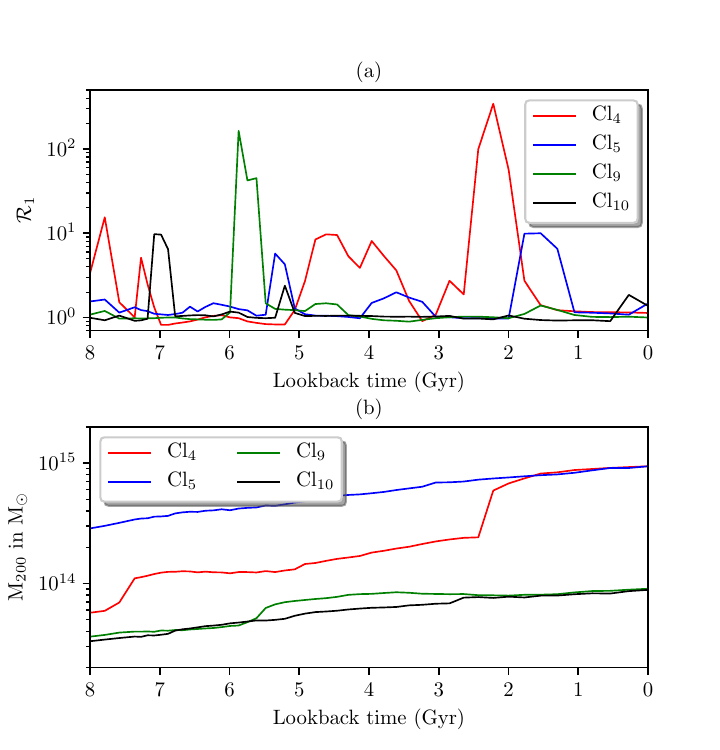}
\caption{{\bf Panel (a):} Time evolution of virial ratio ($\mathcal{R}_1$) of merging and non-merging clusters. {\bf Panel (b):} Mass evolution of the same clusters.}\label{vir-ratio}
\end{figure}

The ratio should be unity for a perfect virialised system. We plot the evolution of this ratio for  pairs of clusters of similar mass but different dynamical state, namely ($Cl_4$ and $Cl_5$) and ($Cl_9$ and $Cl_{10}$) in Figure~\ref{vir-ratio}. Here, we have also shown single ($Cl_4$) and multiple ($Cl_9)$ merging systems. From a comparison of the two panels of Figure~\ref{vir-ratio} one can clearly see that, during mergers (visible as a steep mass increase in Fig.~\ref{vir-ratio}(b)), the virial ratio has boosts (Fig.~\ref{vir-ratio}(a)) too, because the system gains potential energy very fast and also dissipates to various other energy forms in about a Gyr time. In contrast to merger, non-merging systems (i.e. $Cl_5$ and $Cl_{10}$) mostly remained virialised for the whole life span. Fig~\ref{vir-ratio}, clearly shows that a moderate change in the mass is hugely reflected in virial ratio ($\mathcal{R}_1$), making it clearly identifiable. From a qualitative analysis of Figure~\ref{vir-ratio}(a) one can see that, for merging clusters, about one third of the life span remain out of virialisation.

The systems with $\mathcal{ R}_1$ values much larger or smaller than 1 are respectively either in rapid mass accretion phase or in a phase of relaxation. So, the dynamical state of any system can be determined very accurately, if we know how far a system is from $\mathcal{ R}_1=1$.  So, it would be interesting to characterise the dynamical state of the clusters in our sample, in terms of the virialisation parameter defined in Equation~\ref{vir-res1}.

\begin{figure}
\includegraphics[width=1.1\columnwidth]{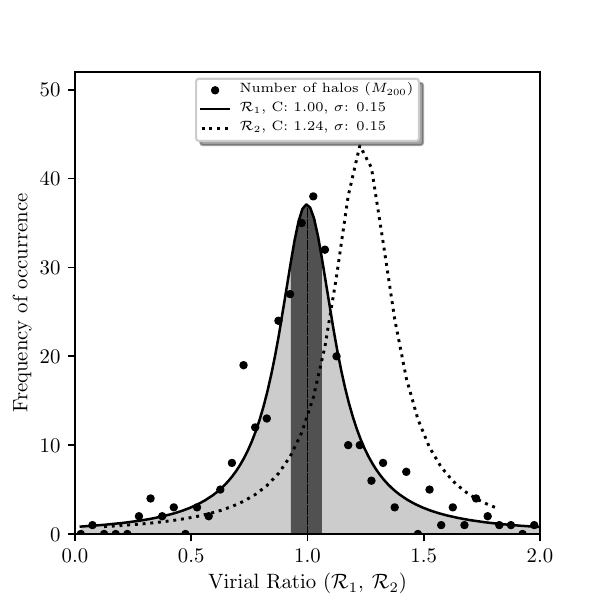}
\caption{ Virial ratio $\mathcal{R}_1$ of the galaxy cluster samples are plotted (solid line) according to their frequency of occurance within bins of size 0.05 in a span of 0-2. A lorentzian is fitted to this frequency plot to compute the statistical mean and the dispersion. Occurence of ratio $\mathcal{R}_2$ has been plotted and fitted as earlier with dotted line.}\label{stat}
\end{figure}

We have chosen about 384 galaxy clusters from different snapshots of our ten realisations. Chosen clusters are in the mass range (virial mass at $M_{200}$) of $5\times10^{13}\; \rm{M_{\odot}}$ to $2 \times 10^{15}\; M_{\odot}$ in between redshift $z=0.5$ to $z=0$. Our simulations though show a statistical mean of this ratio as 0.996, considering a Lorentzian distribution to accommodate the slight asymmetry in the distribution (see Fig~\ref{stat}). For the same systems, if we compute the ratio $\mathcal{R}_2= PE/(2\times KE)$, the statistical mean comes at 1.23 (Fig~\ref{stat}, dotted line), more than 20\% away from $\mathcal{ R}_1=1$. This deviation is known to occur if the surface pressure term arising due to the kinetic stresses at the surface of the collapsing clouds and the external potential due to mass outside the virial radius are not considered for calculating virial ratio \citep{Ballesteros_2006MNRAS,Shaw2006ApJ}. 

\citet{2013MNRAS.436..275W} found that, in their sample, there are 28\% of relaxed clusters. Motivated by analogy with observations, we see that in our sample has about the same fraction of clusters within 10\% from 1.0. Thus we define these objects as ``virialised''. The clusters that are either having low virial ratio or high ratio will be called as non-virialised clusters in our paper. We have found 278 objects i.e. about 2/3 of the all considered galaxy clusters as non-virialised. A high number of non virialised objects could have been resulted due to inclusion of low mass systems (i.e. less than $10^{14}\; \rm{M_{\odot}}$) as smaller clusters are expected to be unstable to dynamical activity or mostly non-virialised \citep{Diaferio1993AJ,Surajit2015,Paul_2017MNRAS}. Non-virialised objects are further divided in to two categories, namely, `HighPE' i.e. objects with higher potential energy (i.e. PE greater than $2\times KE$) and the objects with higher kinetic energy as `HighKE' (i.e. $2\times KE$ greater than PE).

\subsection{Cluster scaling test of the samples}

After categorising our sample, we have computed the virial mass (total) and temperature for all virialised clusters. Assuming that for virialised clusters the assumption of hydrostatic equilibrium holds, we aim to check to which extent the well-known scaling relations are valid for them. Self similar relations between the mass, temperature and the virial radius of cluster are given by $ r_{vir} \propto M_{vir}^{1/3}$ and $T_{vir} \propto M_{vir}/r_{vir}$ i.e. $T \propto r_{vir}^2$ and $T\propto M_{vir}^{2/3}$ (e.g., \citet{Peebles_1980lssu.book,Kaiser_1986MNRAS}). However, observation does not favour this, rather it shows a steeper slope \citep{Maughan_2012MNRAS}. Pre-heating of ICM, non-uniform shock heating, loss of energy to star-formation, non-thermal emissions and so on. i.e. all non-hydrostatic parameters are the major reason for deviation from self similarity \citep{Finoguenov2001A&A,Nevalainen_2000ApJ,Ponman_1999Natur}. Also, as these are connected to the dynamical activity of the clusters, they are the prime reasons for many clusters being non-virialised \citep{Aguerri_2010A&A}. Since deviation from cluster scaling are mainly attributed to dynamical activities, we expect to find the $M-T$ relation to be close to the theoretical value for perfectly virialised objects, but total sample should show a value close to observed value. 

\begin{figure}
\includegraphics[width=1.1\columnwidth]{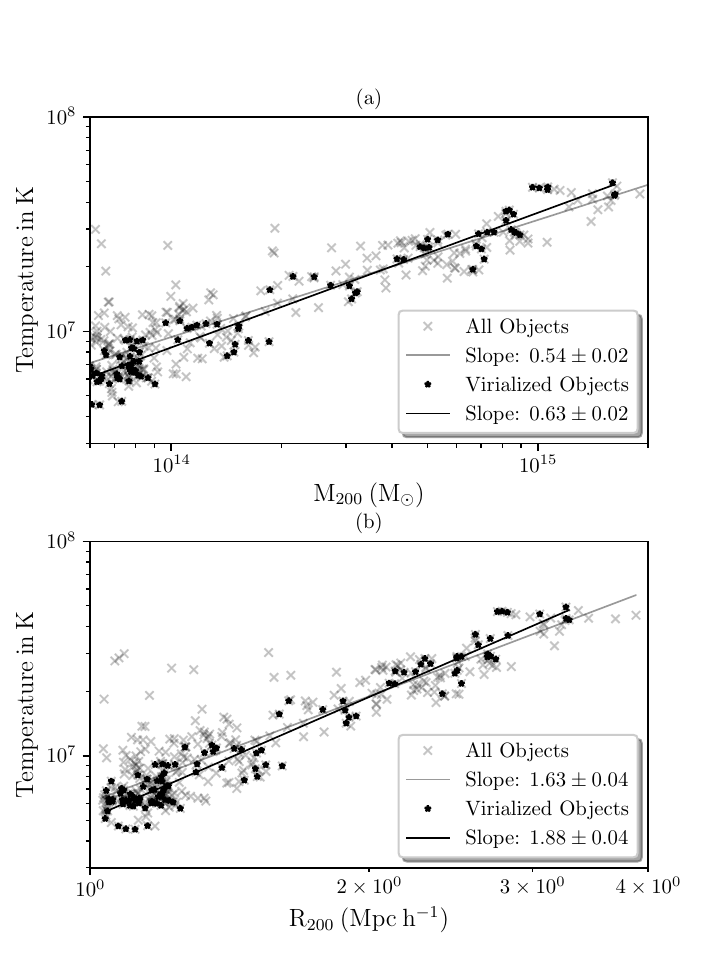}\\
\caption{{\bf Panel (a):} Temperature of the virialised fraction of our sample is plotted against the total mass of the respective clusters. {\bf Panel (b):} Temperature is plotted against the virial radius of the systems. }\label{t-r200-scaling}
\end{figure}

A simple check would help us to validate our choice of virialised objects by testing its $M-T$ relation. To do so,  we have plotted these parameters and fitted scaling laws (see Fig.~\ref{t-r200-scaling}). The statistical fit of the parameters of the chosen virialised clusters in our simulations (Fig~\ref{t-r200-scaling}, black dots and fitted line) shown to obey almost the expected theoretical value of $T \propto r_{vir}^2$ and $T \propto M^{2/3}_{vir}$. The scattered points found in the plots are indicating slight deviation from the exact virial condition. Whereas, when we take all the objects from our sample list (Fig~\ref{t-r200-scaling}, all points and grey fitted line), it closely matches with the observational values of $T \propto M^{0.55}$ and $T \propto r_{vir}^{1.7}$ \citep{Finoguenov2001A&A,Shimizu2003ApJ,Arnaud_2005A&A,Vikhlinin_1999ApJ}. Other scaling laws such as $L_X-T$, $L_X-M$ etc. are presented and discussed in \citet{Paul_2017MNRAS}.\\

\section{Understanding the dynamical states of the simulated systems}\label{dynamical-states-ev}

\subsection{Kinetic and potential energy and the dynamical states of the systems}\label{virial-evol}
 
 \begin{figure*}
\centering
\includegraphics[width=19cm]{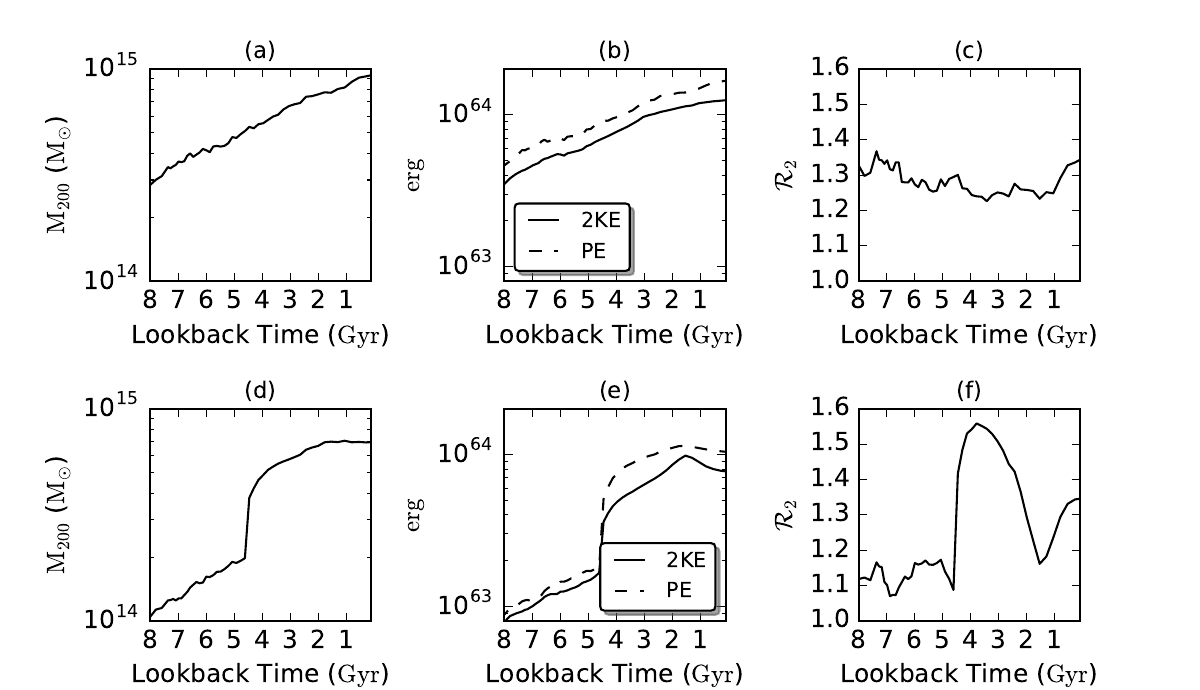}\\
  \caption{Time evolution of mass, potential energy (PE) and twice the kinetic energy (2KE), and ratio $\mathcal{R}_2$ (panel (c)) have been plotted in the three panels (panel (a), (b) and (c)) respectively for a relaxed Cluster ($Cl_5$ of Table~\ref{table_cluster}). Same quantities are plotted (in panel (d), (e) and (f)) for a prominent single merging system ($Cl_7$ of of Table~\ref{table_cluster}).}
 \label{fig:energy-ev}
\end{figure*}

In this section we will further elaborate on the difference between merging and non merging clusters, by studying the time evolution of kinetic and potential energy during the cluster history. We base our analysis on a representative cluster of each category ($Cl_5$ and $Cl_7$ for the non-merging and merging sample , respectively), although our considerations apply to the whole group in general. In Figure~\ref{fig:energy-ev} one can see that, during the whole time span and for both representative clusters, the potential energy is larger than the kinetic component. In the upper row of the Figure~\ref{fig:energy-ev}, evolution of different quantities is plotted for a cluster ($Cl_5$ in our Table~\ref{table_cluster}) that has not experienced any prominent merger during the last 8 Gyr of its evolution. Though the Figure~\ref{fig:energy-ev}(a) and Figure~\ref{fig:energy-ev}(b) show that mass and energies of the system have grown almost 3 times during this period, the ratio of energies (i.e. $\mathcal{R}_2$) has not changed significantly, rather it remained almost constant (Fig.~\ref{fig:energy-ev}(c)). This indicates that a relaxed system, when accreting mass smoothly or through minor mergers, gets enough time to adjust the energy fraction to hold its hydrostatic equilibrium. In the lower row of Figure\ref{fig:energy-ev}, we have plotted the time evolution of the same quantities for $Cl_7$, a cluster that has gone through one significant merger during the same evolution time of 8 Gyr. Figure~\ref{fig:energy-ev}(d) shows a sudden tripling of mass and subsequent slower mass accretion onto the newly formed system which resulted in rapid increment of potential energy within half a Gyr time (in Fig.~\ref{fig:energy-ev}(e)). The growth of the kinetic energy has some time delay with respect to the potential energy. Energy ratio rises very fast, but rate of fall in the energy ratio is much slower and the system takes about 2 Gyr to come back to its pre-merger state (see Fig.~\ref{fig:energy-ev}(f)).  

\begin{figure}
\includegraphics[width=1.1\columnwidth]{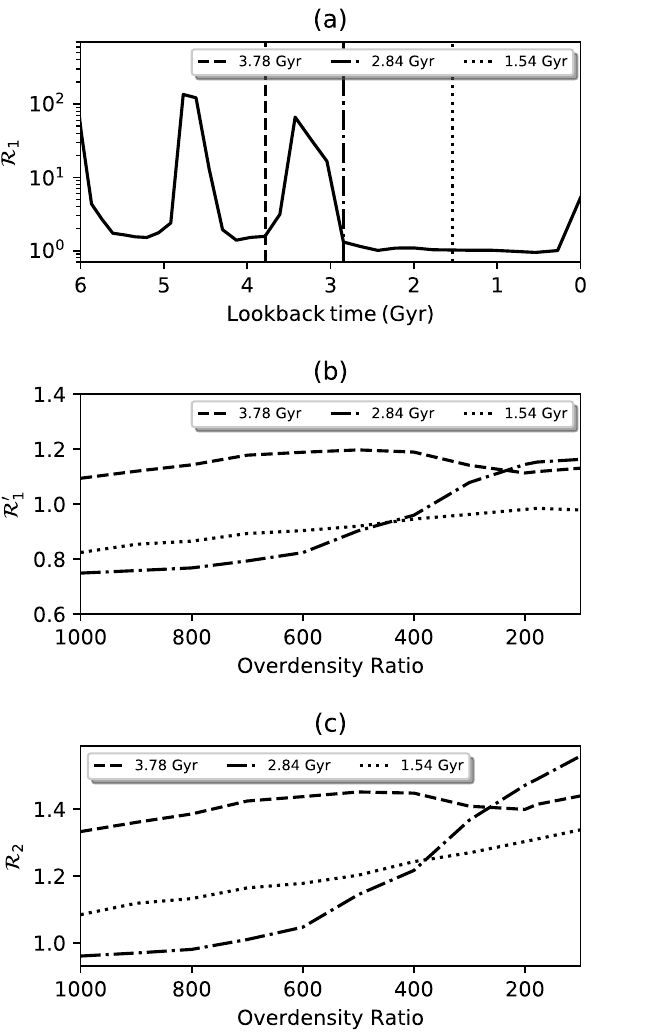}
\caption{ {\bf Panel (a):} Time evolution of virial ratio $\mathcal{R}_1$ for cluster $Cl_2$. Specific dynamical states are indicated on it by dashed line (just after merger), dot-dashed (post merger, KE dominant phase) and dotted line (relaxed phase). {\bf Panel (b):} Radial profile of $\mathcal{R}'_1$ at the specific phases marked in panel (a). {\bf Panel (c):} Profile of $\mathcal{R}_2$ against radial overdensity ratios for the same states.}\label{radial-en-ratio}
\end{figure}

Further, to study the spatial evolution of different energies within a cluster at different dynamical conditions, the radial profile of virial ratio has been plotted (see Fig~\ref{radial-en-ratio}). For this study, our virial ratio definition has been modified slightly from what is defined in Section~\ref{dyna-stat}. External potential by definition is the potential due to the mass outside virial radius i.e., mass residing between $r_{200}$ to at most the second turn around radius. So, inside the virial radius, the quantity becomes ill defined and show spurious results. Considering this fact, for radial profiles we have modified the virial ratio definition as

\begin{equation}
\mathcal{R}'_1 = \frac{(U_{int} -E_s)}{2\times KE}
\end{equation}

To make it easier to understand, in Figs.~\ref{radial-en-ratio}(b) and (c) we plot the over-density ratio on the $x$-axis. Virial radius in this scale would be plotted as $r_{200}$. Use of this scale makes it easier as it automatically normalises the $x$ axis for comparing a cluster at different times of its evolution. The radial profiles of this quantity $\mathcal{R}'_1$ have been plotted in Fig.~\ref{radial-en-ratio}(b) at selected times during the evolution of the cluster $Cl_2$. Each of these times have been chosen to correspond to specific phases during the merger and subsequent relaxation, as visualised and marked in the evolution of $\mathcal{R}_1$ (Fig.~\ref{radial-en-ratio}(a)). A merger event in $Cl_2$ has started at look back time of $t=3.78$ Gyr which is depicted by dashed vertical lines. The next one shows the time at $t=2.84$ Gyr when potential energy gained due to mergers has mostly been converted to kinetic energy i.e. highest kinetic energy phase and shown as dot-dash line. Finally, at 1.54 Gyr, when the cluster has again come to relaxed condition, has been shown as dotted line.

Fig~\ref{radial-en-ratio}(b) shows a possible deviation from the  well known spherical accretion model. The cluster core in our simulations never seem to remain in virial equilibrium. Though in relaxed phase, virial ratio $\mathcal{R}'_1$ makes up to almost 1 at $r_{200}$, both merging and relaxing phases show a significant deviation from unity. As we go inside the cluster, the merging state is dominated by potential energy, while the relaxing state is mostly dominated by kinetic energy (see Fig.~\ref{radial-en-ratio}(c)). But only in case of relaxed condition, maximum range of $x$-axis of the cluster till $r_{200}$ has values near to unity. 

\subsection{Shocking the cluster medium}\label{shock-mach}

In this Section we study the shock properties inside the clusters, mainly the representative parameters of shocks such as shock strength and area that determines the CR emission in the cluster medium. For this, we considered the region inside the virial radius (i.e. r$_{200}$). Within this cluster volume, shocks have been identified with means of unsplit velocity jumps across the cells \citep{Vazza_2009MNRAS,Vazza_2011MNRAS,Skillman2008ApJ}.

In the simulated clusters, in general the shock Mach numbers are in the range $\mathcal{M}=1-2$ in the core region i.e $r_{1000}$. The Mach number goes to almost $\mathcal{M}=4$ in regions beyond the core but inside virial radius (i.e. $r_{200}$) (Fig.~\ref{fig:mach-CRF}, panel 1, is shown as a representative map). It rarely reaches as high as $\mathcal{M}=10$. Outside the radius (i.e.  $r_{200}$) it goes beyond $\mathcal{M}=10$. The shocks inside the virial radius are usually called the internal shocks \citep{Miniati2000ApJ,Skillman2008ApJ}.

\begin{figure}
\includegraphics[width=1\columnwidth]{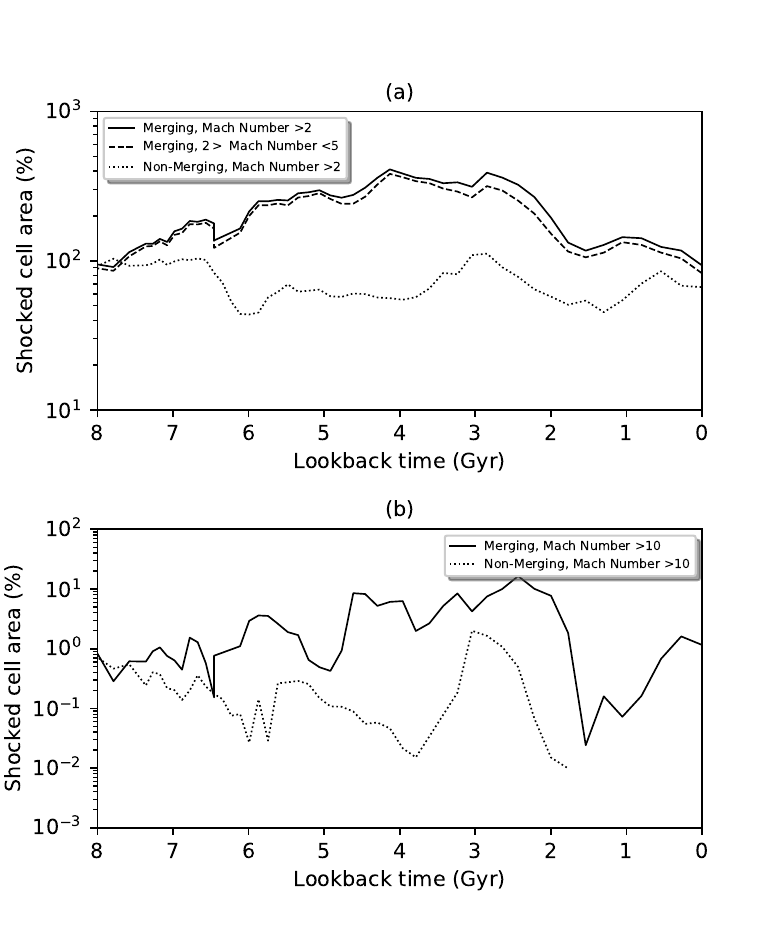}
\caption{{\bf Panel (a):} Time evolution of percentage of shocked cell surface area (of Mach number $\mathcal{M}>2$ (solid line) and $2<\mathcal{M}<5$ (dashed line) for the merging cluster ($Cl_2$) and $\mathcal{M}>2$ (doted line)) for non-merging cluster ($Cl_5$). The values are normalised to the virial ($r_{200}$) surface area of the clusters at that point of time. {\bf Panel (b)} Shocked cell area for $M>10$ for the same merging (solid line) and non-merging (dotted line) systems as above respectively.}\label{mach-sv-sa}
\end{figure}

Numerical shocks in grid-based schemes are typically smeared over 3 cells in the direction of shock normal. The shocked surface area has been approximated as the sum of the face area of the cells tagged as shocked (See \citet{Vazza_2017MNRAS}).
We have further plotted the percentage of area of shocked cell surfaces (SCS) to the total surface area at $r_{200}$ for the cluster Cl$_2$ for Mach numbers above $\mathcal{M}>2$ and for the last 8 Gyr of its evolution (see Fig.~\ref{mach-sv-sa}). It is also seen in the merging clusters that most of the shocked cell area is actually occupied by the shocks with Mach number between $\mathcal{M}=2-5$ (dashed line in Fig.~\ref{mach-sv-sa}(a)). This plot clearly reveals that during merger, SCS area occupies more than 200$\%$  and reaches as much as \footnotemark 400\%  in certain most active phases of the total virial surface area of the cluster, with a sharp contrast to the usual SCS of 100$\%$ percent or less in the cluster in relaxed condition (Fig.~\ref{mach-sv-sa}(a), dotted line). Another important observation is that, at low redshift, the SCS rapidly falls down by about an order of magnitude. In a relaxed system, Mach number beyond $\mathcal{M}=10$ is hardly reached, whereas, during mergers, surface of strong shocked cells sometimes increase by about $10^2$ folds than that of a relaxed system (see Fig.~\ref{mach-sv-sa}(b)). 

\footnotetext{Since luminosity has been calculated from all shocked cells emerged inside the cluster virial volume and there can be many shocks (as seen in Fig.~\ref{fig:mach-CRF}, panel 1) filling most part of the volume with multiple layers, we can get more than the surface area of the cluster as shocked. }

\section{Dynamical states and its connection to energy evolution}\label{res-dic}

In this section, we combine both the defined diagnostics for the dynamical states of clusters and the proxies to observable quantities to get an overall impression on their correlations, during the cluster evolution . We begin with the CR production due to the acceleration of thermal particles by DSA mechanism.
 
\subsection{Energetics of evolving dynamical states of galaxy clusters}\label{dyn-phase}

\begin{figure*}
\includegraphics[width=1\textwidth]{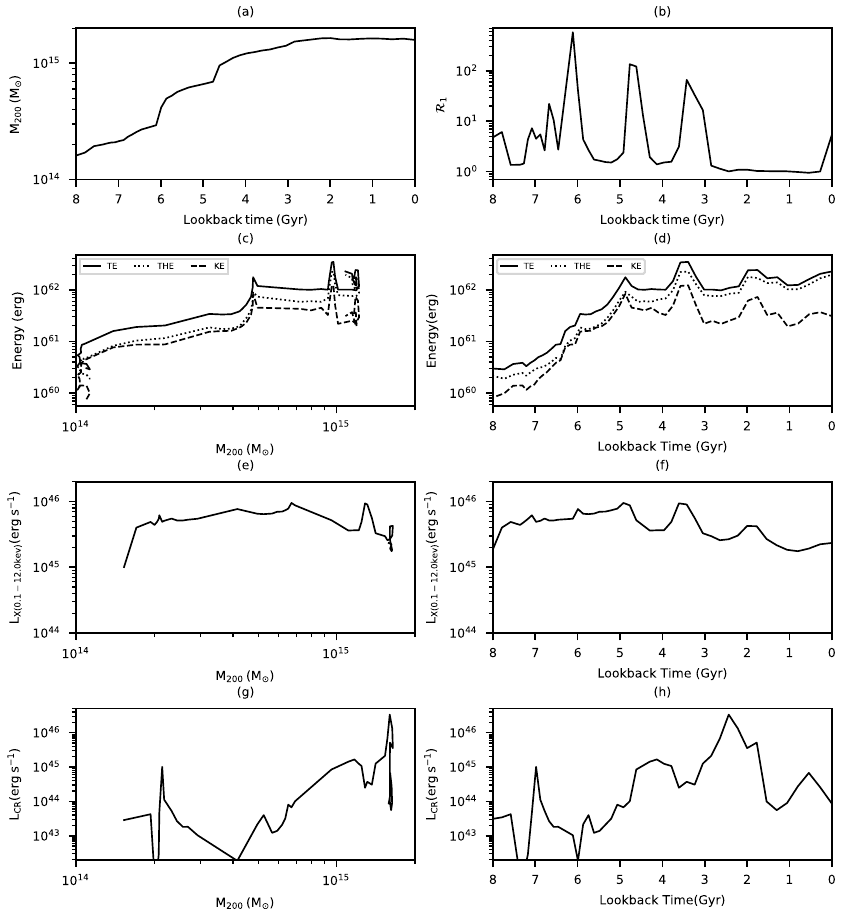}
\caption{Eight-panel plot of our representative merging cluster ($Cl_2$ of Fig.~\ref{sample}). Mass and $\mathcal{R}_1$ has been plotted with lookback time in panels (a) and (b) respectively. Total, thermal and kinetic energy have been plotted against $M_{200}$(c) and Lookback time (d). Further, X-ray luminosity and CR luminosity evolution shown against $M_{200}$ (panels (e) and (g) and lookback time (panels (f) and (h)).}\label{merg-phase}   
\end{figure*}

In our simulations, the main representative merging cluster i.e. $Cl_2$ has gone through multiple mergers during its 8 Gyr of evolution time as shown using the total mass plot in Figure~\ref{merg-phase}(a). This clearly indicates three major mergers roughly at look-back time of 6, 4.5 and 3 Gyr with merging mass ratio varying in the range of 1:1 to 1:3. Virial ratio $\mathcal{R}_1$ (see Section~\ref{dyna-stat}) which represents the dynamical state of the system (see Section~\ref{dyna-stat}~and~\ref{virial-evol}) peaks at these points indicating a significant change in dynamics of the system (see Fig.~\ref{merg-phase}(b)). On each merger, the total energy (computed as the sum of radial averaged quantities) of the system (thermal energy plus kinetic energy of baryons) increases substantially (see Fig.~\ref{merg-phase}(d)). It is important to stress that the total energy of the system peaks almost a Gyr after each merger as indicated in mass (Fig.~\ref{merg-phase}(a)) and in the virial ratio (Fig.~\ref{merg-phase}(b)). At each instance the time delays of energy are almost the same (i.e. about one Gyr); at about look-back time 5, 3.5 and 2 Gyr respectively (Fig.~\ref{merg-phase}(d)). These are at the same point of time when virial ratio has become lowest after each merger (Fig.~\ref{merg-phase}(b)). The baryon kinetic energy also peaks almost at the same time. This is consistent with theory as these are the phases where most of the potential energy is supposed to be converted to kinetic energy. It is also noticed that relative increment in gas thermal energy is slightly more than the relative increment in total energy. This is due to the gas gaining energy by interaction of baryon particles as well as the additional PdV work by  the shocks \citep{Sarazin2001cghr}. The intracluster medium dissipates a fraction of this energy through different radiation processes within less than a Gyr. Through bremsstrahlung X-rays emission, the system loses its thermal energy while the kinetic energy gets used up to produce cosmic rays by shock acceleration of charged particles. Our results (Fig.~\ref{merg-phase}(f)) show that the phase of X-ray luminosity gain coincides with the gas thermal energy gain due to mergers. Though the cluster starts emitting X-rays in plenty almost a Gyr year after the mergers, cosmic rays take more than one and half Gyr year to reach the peak (Fig~\ref{merg-phase}(h)). The time delays of X-ray and CR injection peaks are consistent in all three mergers for $Cl_2$ as well as in all other merging clusters (Table~\ref{table_cluster}). The brightest phase of CR injection occurs about 1.5 Gyr after the merger has taken place.

Though the total energy of the system due to mergers increases only by few times in magnitude, the CR injection energy shoots up to almost two orders in magnitude. CR injection and X-ray luminosity both decrease at the low redshift, although the total energy of the cluster keeps growing (see Fig.~\ref{merg-phase}(f) and Fig.~\ref{merg-phase}(h)). The outcome is also supported by the fact that at lower redshift shocked cell area consistently decreases (see Fig~\ref{mach-sv-sa}(a)) reducing both shock heating and CR acceleration efficiency. It is also noticed that percentage of fall of total CR luminosity is more compared to x-ray luminosity (see Figure~\ref{merg-phase}, panel~(f),(h)). From the figure~\ref{merg-phase}(e),(g) it can be seen that though merger happens due to rapid accumulation of mass, X-ray emission and CR injection are not correlated with mass.

\subsection{Merging state and its connection to cluster energetics}\label{comp-M-noM}

\begin{figure*}
\includegraphics[width=1\textwidth]{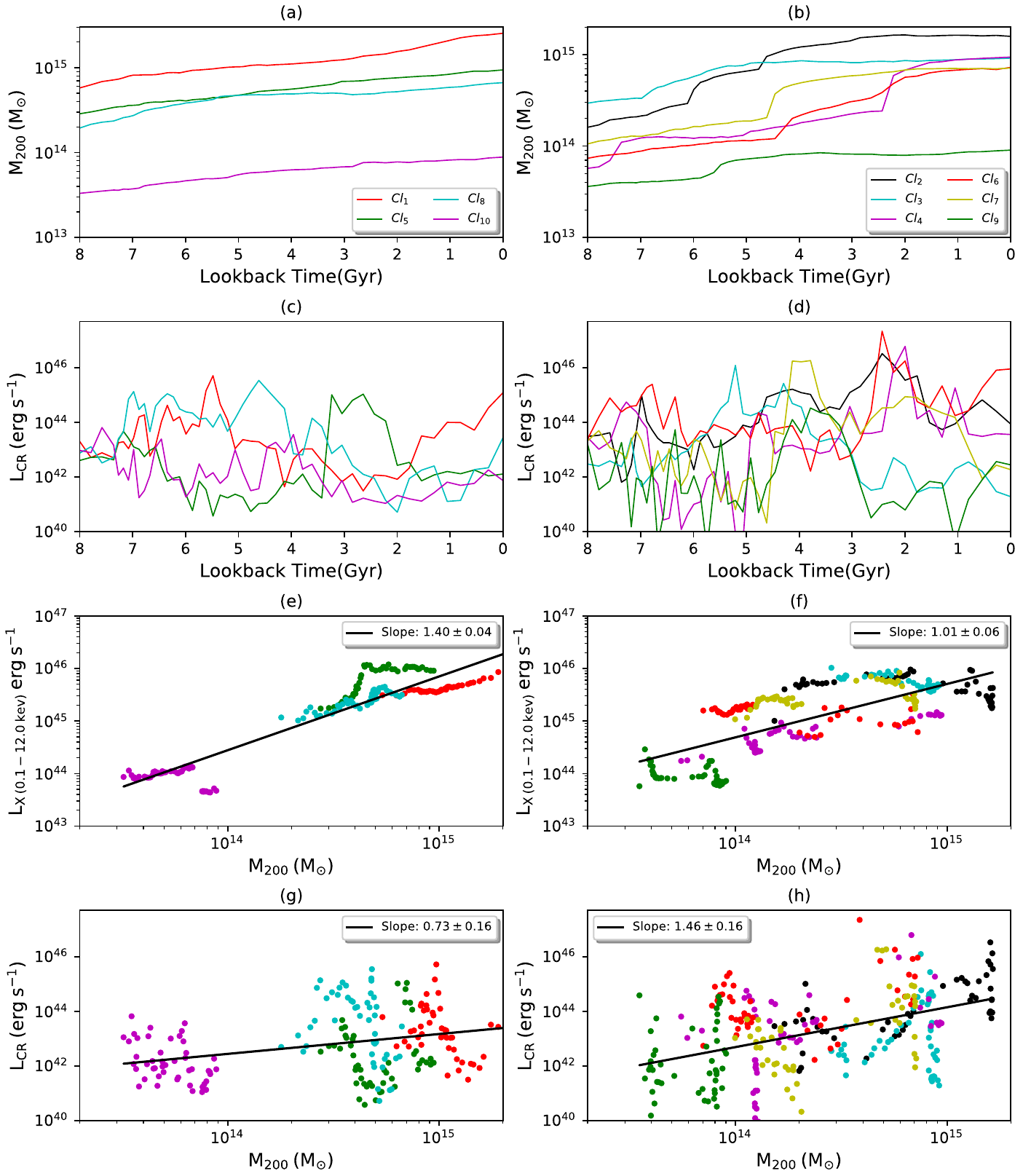}
\caption{Evolution of mass and CR injection flux has been plotted against lookback time in panels (a) and (c) and X-ray luminosity and CR luminosity has been plotted against virial mass of the non-merging clusters in the panel (e) and (g). The same variables are shown for merging clusters in panels (b),(d),(f) and (h) respectively. Colours represent each cluster as given in panels (a) and (b).} \label{merg-no-merg}
\end{figure*}

In Figure~\ref{merg-no-merg} we compare the mass accretion history and the luminosities in X-rays and CRs of merging and non-merging systems (see Section~\ref{mer-non-mer}). Different colours represent different clusters. We have observed that a single significant merger can alter the energy distribution of the system and can dominate for over 2 Gyr (see Fig~\ref{fig:energy-ev}). A major merger can induce turbulence in the system that persists for more than 2 Gyr and up to 4 Gyr in case of core oscillations \citep{Paul2011ApJ} i.e. almost half of its evolution time scale. Cluster mergers are also capable of altering energy budget up to few virial radii from the centre \citep{Iapichino_2017MNRAS}.  If this point is considered in terms of cluster scaling laws, it brings to the suggestion that a system once merged may permanently deviate from the scaling that a non-merging system would follow \citep{Ascasibar2006MNRAS,Krause2012MNRAS}. Scaling laws in X-rays (thermal) and CRs (non-thermal) would then indicate it.

In Figure~\ref{merg-no-merg}, the left column shows different features of non-merging systems and the right column shows that of merging systems. In Figure~\ref{merg-no-merg}(a), four non-merging systems have been plotted that show a smooth increment of mass during an evolution time of 8 Gyr. On the other hand Figure~\ref{merg-no-merg}(b) shows six merging clusters having one or multiple jumps in total mass, indicating that a merger has occurred in the system. In Figure~\ref{merg-no-merg}(c) and (d), we have plotted the corresponding CR emission luminosity. Though the time evolution of mass looks rather smooth outside merger events, so is not for the evolution of $L_{CR}$. The evolution of this quantity shows numerous fluctuations with time, pointing to the extreme sensitivity of the CR acceleration to even minor variations of the shock location and strength in the ICM. The cosmic ray emission shows a jump of several orders of magnitude each time a cluster goes through a significant merger. This  behaviour can be observed for all the merging systems as plotted in panel (d). The finding described in the Section~\ref{dyn-phase} found stronger support in Figure~\ref{merg-no-merg}(b) and (d), where it can also be noticed that the major CR production peaks come approximately after 1.5 Gyr from every significant mass jump i.e. the mergers. While looking at a large number of systems in our sample, we have noticed (in Fig.~\ref{merg-no-merg}(g) and (h)) that the clusters with same masses, the average CR luminosity of a non-merging system is almost an order of magnitude less than that of a merging system (compare Fig.~\ref{merg-no-merg}(c) and (d)). From the same figure, a key conclusion can be drawn about the non-efficient CR acceleration in the high mass systems. It shows that though the initial mergers bring up the injected CR population to several orders of magnitude higher than its value at relaxed condition, injected CR population falls to the level of almost same as the non-merging systems in the final stage of its evolution. For a possible explanation of this feature, we speculate that a cluster that goes through earlier mergers is hotter than the usual clusters. This hot medium will reduce the Mach number of the internal shocks and therefore the shocked surface area and as a consequence DSA becomes less effective.\\

\subsubsection{Scaling relations for merging and non-merging systems}\label{scale-M-no-M}

In figure~\ref{merg-no-merg}(e) to (h), we show the evolution of X-ray and CR luminosity in six merging clusters and four non-merging systems. The points in the plots with same colour represent the same cluster at different times during its evolution. The comparison of the resulting scaling laws $L_{CR} \propto M^{\alpha}$ and $L_X \propto M^{\beta}$ show a clear difference between the merging and non-merging systems. While merging systems show a steeper slope $\alpha =$ 1.46 for CR luminosity with scatter value of $\sigma$ = 0.16, non-merging systems are following much flatter slope of $\alpha =$ 0.73 with scatter value of $\sigma$ = 0.16. The situation got just reversed in case of X-ray emission where merging clusters have much flatter slope of $\beta =$ 1.02 (with $\sigma$ = 0.06) than the non-merging one which is $\beta =$ 1.40 (with $\sigma$ = 0.04). This shows that merging significantly changes the energetics of the galaxy clusters. Moreover, a much higher dispersion value found in CR luminosities compared to X-rays is indicative of a transient nature of CR injection connected to cluster dynamics.

 \subsection{Virialisation and its connection to cluster energetics}\label{virial-connects}
 
 In Section~\ref{dyna-stat} and \ref{virial-evol}, other than the mergers, we have devised another method of determining the dynamical phase of galaxy clusters using virialisation as yardstick. Figure~\ref{stat}~and~\ref{radial-en-ratio} clearly indicate a difference in the dynamics of the clusters according to its state of virialisation. As discussed in the last paragraph of Section~\ref{dyna-stat},  we have classified systems as virialised and non-virialised and further the non-virialised systems to HighPE and HighKE. We have already discussed in Section~\ref{virial-evol}~and~\ref{dyn-phase} about how this categorisation helped us in understanding the energy evolution in individual clusters. Further, we will try to use this method for distinguishing objects using the large sample set.
   
 \begin{figure*}
\includegraphics[width=1\textwidth]{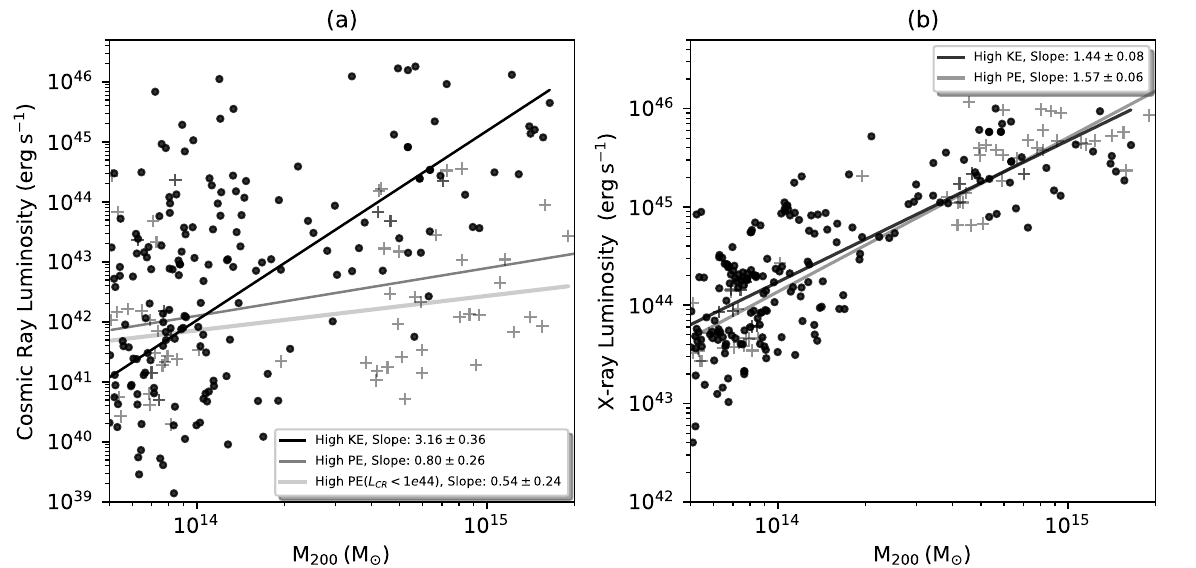}
\caption{ CR (panel (a)) and X-ray luminosity (panel (b)) of HighKE (dark dots) and HighPE (grey pluses) are plotted against the mass ($M_{200}$).}\label{Energy-dyn-stat2}
\end{figure*}

We have noticed that the clusters in merging state gain high potential energy very rapidly. But significant heating of the medium along with increment in kinetic energy happens in the post-merger phase when the medium tries to attain equilibrium through generation of shocks, a phenomenon called as `violent relaxation' \citep{Wise_2007ApJ}. Heating continues till the cluster again comes to relaxed state. This can be understood from the thermal X-ray emission shown in Figure~\ref{merg-phase}(f) and  in a large sample in Figure~\ref{merg-no-merg}(f). Further, CR acceleration, as by-product of shocks, sets in only after a Gyr from the merger i.e. in the post-merger phase. Since the central part of the cluster is the hottest, Mach number of the shock is usually low there i.e. below $\mathcal{M}=2$ (within $r_{1000}$). Shock becomes stronger beyond the cluster core where the medium is relatively colder. Most of the shocks with Mach number $\mathcal{M}=2-5$ are found between $r_{1000}$ and the virial radius of the cluster (see Fig~\ref{fig:mach-CRF}(a)). This tells us that the thermalisation and particle acceleration actually takes place only in the post-merger phase (i.e. HighKE) indicating the introduction of a significant change in energy distribution in the ICM at post-merger compared to the virialised and to the merger phases (i.e. HighPE).
  
 \subsubsection{Scaling relations from the state of virialisation}\label{scale}

Since dynamical activity in the galaxy clusters modifies the energy budget of the clusters, the contribution to luminosity from the thermal and non-thermal components are expected to change during the cluster evolution. In Figure~\ref{Energy-dyn-stat2}, we have plotted X-ray and CR luminosity of HighKE and HighPE objects against mass. Figure~\ref{Energy-dyn-stat2}(a) shows that highKE clusters have very steep scaling of cosmic ray luminosity with mass i.e. $L_{CR} \propto M^{\alpha_{*}}$ with $\alpha_{hK}=3.16\pm\sigma = 0.36$. The scaling exponent of the HighPE clusters is much flatter, namely $\alpha_{hP} =0.80\pm0.26$. It can be noticed that most of the HighPE objects ($M> 5\times 10^{14}M_{\odot}$) have $L_{CR}$ at least an order of magnitude lower than the HighKE objects, therefore, removing few outliers ($L_{CR}> 10^{44}$ erg s$^{-1}$) the slope becomes even flatter ($\alpha_{hK}=0.54\pm0.24$). X-ray emission from HighKE has a scaling relation of $L_X \propto M^{\beta_{*}}$ with $\beta_{hK}=1.44\pm0.08$ which does not show much difference with the HighPE objects ($\beta_{hP}=1.57\pm0.06$), but the marginal change of the scaling slope is in the opposite direction to CRs which is in well agreement with the finding in Section~\ref{scale-M-no-M}. Importantly, it can also be noticed from these slopes that the average CR energy is less than $10$\% of the X-ray energy in all mass ranges for HighPE objects (see Figure~\ref{Energy-dyn-stat2}(a) and (b)). However, objects at the high mass ($M\ge 5\times 10^{14} M_{\odot}$), in the HighKE group have larger $L_{CR}$ than the ones of the HighPE group. The difference is more than an order of magnitude, and HighKE clusters have $L_{CR}$ almost as large as their $L_X$. This becomes an indicator of post-merger and non-virialised systems for the high mass objects.

  \subsection{Mach number distribution and its connection to CR emissions}\label{mach-CR}
  
\begin{figure}
\includegraphics[width=1.1\columnwidth]{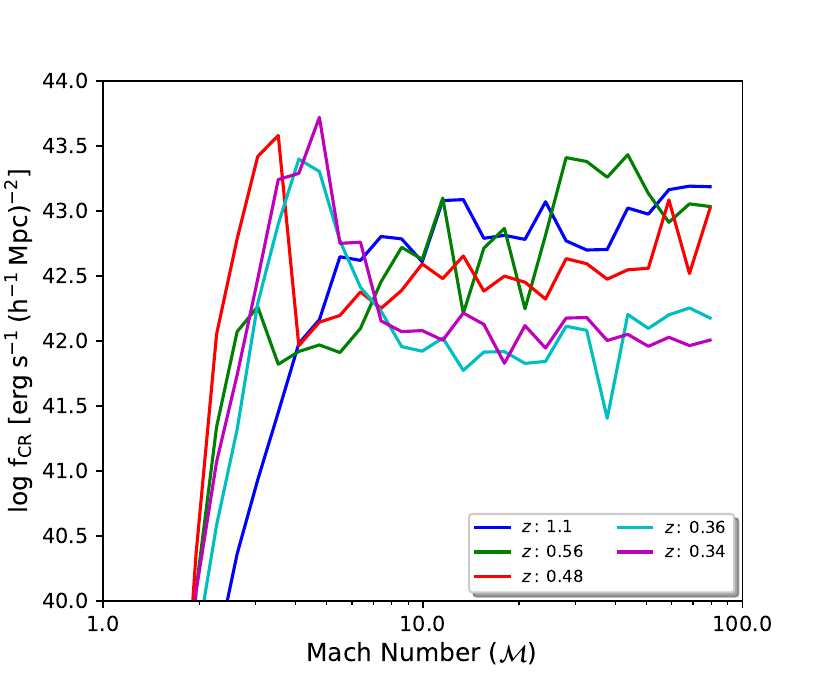}
\caption{Cosmic ray flux has been plotted against Mach number at different redshifts for the cluster Cl$_2$.}\label{mach-cr}
\end{figure}

Cosmic ray injection due to particle acceleration is usually known to be a steep function of shock strength \citep{Ryu_2003ApJ}. In Figure~\ref{mach-cr}, the evolution of the CR flux as a function of the shock Mach number for the representative merging cluster $Cl_2$ is shown. The plot clearly shows that the most CR acceleration is happening for the Mach number ranging $\mathcal{M}=2.5-5.5$ (peak at about $\mathcal{M}=3.5$) during the evolution of the merging clusters, in agreement with the work of \citet{HONG2014ApJ}. We have also found that the peak of cosmic ray emission is seen to move towards higher number starting from $\mathcal{M}=3-4.5$ as the time passes after a merger. This happens on a time scale of $2-3$ Gyr during each mergers. Similar plots as Figure~\ref{mach-cr} for other merging clusters, not shown here, demonstrate that the range and the peak slightly varies in different merging clusters in our sample but the basic trend holds for all mergers. This also indicates that most of the CR acceleration is probably happening due to the main merger shock (see Section~\ref{shock-mach}), which increases its Mach number while propagating in a colder medium towards the cluster outskirts.

We further notice that this variation of range and the peak of effective Mach number is a steep function of the shock area. Figure~\ref{mach-sv-sa}(a) shows the differential area with Mach number ranges. It is evident from this plot that almost all shocked cells are associated to shocks with Mach number in the range $\mathcal{M} \sim 2-5$, while, Mach number above $\mathcal{M}=10$ is occupying very minor fraction of the virial surface area (about 10\% only, see Fig.~\ref{mach-sv-sa}(b)). Shocked area in $Cl_1$, the non-merging cluster, is seen to be almost flat and stays at a level of 100\% or below throughout the cluster evolution. The same behaviour is also shown by the CR luminosity which has a rather flat evolution with time and stays at a level of 10$^{42-43}$ ergs$^{-1}$ (Fig.~\ref{merg-no-merg}(c)). It has also been noticed that at low redshift, the shocked area in the merging cluster $Cl_2$ has decreased to a value typical of non-merging clusters. As a consequence, also its CR luminosity decreases to a value typical of non-merging objects $\sim$10$^{43}$ ergs$^{-1}$, Fig~\ref{merg-no-merg}(d).

\subsection{Re-acceleration of pre-existing Cosmic Rays}\label{pre-exist}

So far, we discussed about CR production by the acceleration of thermal particles. However, it is well known that clusters host a significant amount of preCRs that may affect the CR injection. Re-acceleration of preCRs is more efficient at low Mach shocks \citep{Kang2013ApJ}, seen in plenty inside the cluster virial radius, and during the initial stage of mergers as shown in Figure~\ref{fig:mach-CRF}(a) in this study, as well as reported in several simulation studies \citep{Miniati_2001ApJ,HONG2014ApJ}. For this reason it is important to discuss the role of the re-acceleration of preCRs in clusters. The acceleration of preCR is very sensitive to the pre-existing CR pressure in the medium. This would need to be properly computed at runtime in the simulation. This is not the case in our approach, however we mimic the presence of preCRs in a test where an initial CR to thermal pressure ratio of 0.05 and a slope of CR proton  s = 4.5 in each of the cells inside the cluster volume are assumed.  Then the piecewise fitting formula from \citet{Kang2013ApJ} is implemented.

\begin{figure*}
\includegraphics[width=18cm]{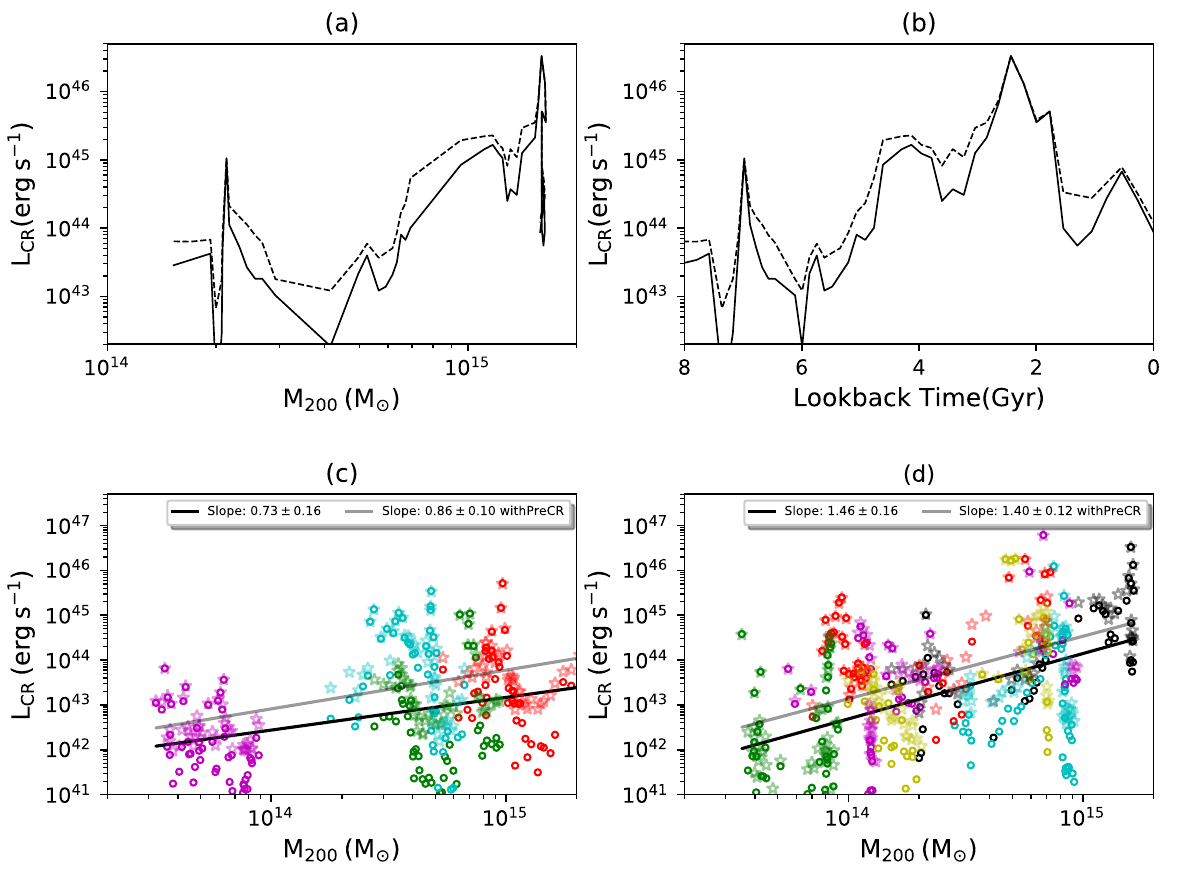}
\caption{Evolution of CR luminosity computed for thermal (solid line) and preCRs are shown against $M_{200}$ (panel~(a)) and lookback time (panel~(b)) for cluster $Cl_2$. Same for all non-merging and merging clusters are plotted in panel~(c)~and~(d) respectively. Colours represent each cluster as given in Fig.~\ref{merg-no-merg}(a)~and~(b) with faint colour stars indicating CR injection from preCRs.}\label{fig:compare-pre-no-pre}
\end{figure*}

In this test, preCRs are found to alter the particle acceleration significantly at low Mach numbers as shown by the comparison of Figure~\ref{fig:mach-effi} and Figure~\ref{fig:Lcr-mach}. CR acceleration efficiency for preCRs peaks at lower Mach shocks i.e. $\mathcal{M}=3$ (Appendix~\ref{appen-preCR-study}) as compared to $\mathcal{M}=3.5$ for thermal particles (Fig~\ref{mach-cr}).  Shocks when emerged and propagated in the cluster central region (i.e. at $< r_{500}$) have a low Mach number and efficiently re-accelerate preCRs during early stages of mergers as seen in a representative slice plot in Figure~\ref{fig:Cr-effi-pre-ext}~and~\ref{fig:PCRbyPTH-pre-ext}. It is also noticed that on an average, the CR pressure is much higher inside $r_{500}$ of the clusters as discussed in Appendix~\ref{appen-preCR-study}, Figure~\ref{fig:PCRbyPTH-pre-ext}. Figure~\ref{fig:compare-pre-no-pre} presents the comparison of a few crucial results obtained for CR acceleration of thermal particles and the re-acceleration of preCRs. With the example of our reference merging cluster ($Cl_2$), we recognise no significant difference in the basic nature of production, temporal and spatial evolution of CRs using these two schemes. However, in Figure~\ref{fig:compare-pre-no-pre}(a)~and~(b), we notice that the level of CR luminosity during non-merging as well as the early period of merger is highly dominated by preCRs, though at the peak, we found no difference. Finally, as indicated by a single system in the first two panels, no significant difference is observed in evolution pattern when several systems are studied (see Figure~\ref{fig:compare-pre-no-pre}(c)~and(d)). However, the average value of CR luminosity combining all these systems seems to have altered significantly with a change of almost half an order of magnitude. Notably, the change is similar in both merging and non-merging systems as shown in Figure~\ref{fig:compare-pre-no-pre}(c)~and~(d).

\subsection{Limitations of this study}\label{limit}

Since DSA is widely accepted to be the main mechanism behind cosmic ray acceleration in the large scale structures \citep{Kang_1997MNRAS,Miniati_2001ApJ}, our study considers only DSA as the particle acceleration mechanism.  In this context, other works motivated by the observed cluster scale radio emissions have also pointed out about the contribution from turbulent re-acceleration \citep{Brunetti_2016PPCF,Brunetti_2011MNRAS,Eckert_2017ApJ}. We are well aware of the important role of AGN activity and SN explosion in the CR acceleration, especially for providing a pristine CR pool for both DSA and re-acceleration model \citep{Kang2007ApJ,Brunetti_2011MNRAS}. 

Primarily, our study is based on the acceleration of thermal particles by DSA. We also present a constraint model of CR acceleration by DSA that considers preCR population \citep{Kang2013ApJ}. But, we did not work at all on turbulent re-acceleration of preCR or their (AGN and SN) direct contribution. Further, our computation of CR acceleration is done using the piece-wise fitting function of CR acceleration efficiency $\eta(\mathcal{M})$ from \citet{Kang2013ApJ} as the post process of the snapshots taken from our hydrodynamics plus N-body simulations. So, our calculations consider only instantaneous CR acceleration or injection, and no time evolution solutions. However, the peak of CR luminosity for merging clusters is, in some cases, one order of magnitude or more larger than the pre-merger value, partly justifying the instantaneous approach for the CR acceleration in those cases. CR transport mechanism has also been ignored here. 

Study with preCR as discussed in Section~\ref{pre-exist} is a tentative study that only shows a possible picture of the effect of re-acceleration of preCRs in overall CR production and evolution scenario. Diagnostic used here is indicative only and prone to over or underestimate the physical parameters, therefore, should not be taken as a generic case while using it as a benchmark to explain any theoretical or observational upper limits.

Since the physical process of CR emission evolves with time and transport of high energy particles alters its energy and distribution, to obtain a more realistic picture one has to implement all the above mentioned physics in the calculations as well as a computation scheme that takes care of time evolution. Therefore, this study is aimed only to understand the CR injection properties i.e. transient nature of CR emission rather than the total accumulation of CRs inside the clusters.

\section{summary}\label{sum}

In this paper, we report an extensive study on the cluster dynamics and energetics as well as origin, injection and temporal and spatial evolution of cosmic rays in merging and non-merging galaxy clusters. We have performed cosmological simulations with the AMR code \textsc{Enzo} \citep{Bryan_2014ApJS}. The cosmic ray production in space and time is elucidated in detail (see Section~\ref{comp-M-noM}) as a post-processing steps of cosmological simulations using the CR acceleration efficiency models of \citet{Kang2013ApJ}. We have primarily considered the acceleration of thermally energised particles by DSA, as well as discussed in brief, a special case of re-acceleration of preCRs, improving over the works of \citet{HONG2014ApJ}. We find a strong connection between the dynamical states of galaxy clusters and the production of CRs. Energy distribution in clusters are shown to be directly related to merger activity, fractional area covered by shocks and on how far a cluster is away from virialisation (according to the definition of virialisation in Section~\ref{dyna-stat}). The main outcomes of this study are the following:\\

\begin{itemize}

\item We show that the percentage of shocked surface area in a cluster is crucial for CR emission efficiency. In most of the clusters, Mach number in the range between  $\mathcal{M}=2-5$ captures the most of the shocked area and is therefore the most effective generator of CRs. Since this is the Mach number range of internal or merger shocks, it can be inferred that merger events are the most effective engines for CR acceleration from clusters. But, the high Mach number ($\mathcal{M}>5$) merger and accretion shocks at cluster periphery are found to be important for deciding the brightest phase of CR injection i.e. emission peaks in the clusters.\\

\item Our study shows that there is a consistent time delay in the peaks of X-ray emission and CR injection, considered to be the tracers of thermal and non-thermal energy production during merger events. The X-ray luminosity and total energy of merging galaxy clusters are observed to attain a maximum after 1 Gyr of the merging event, whereas, luminosity of CR injection reaches its peak after about 1.5 Gyr after the merger. The same pattern is noticed in both the cases of acceleration of thermal as well as preCR particles. This non-concurrence of the thermal and non-thermal injection peaks found in this study needs further investigation for its impact on the evolution of CR-derived observables from galaxy clusters.\\

\item Clusters with merging history permanently deviates from the mass scaling laws for X-ray and CR injection luminosity of  non-merging systems (see Section~\ref{scale-M-no-M} and Figure~\ref{merg-no-merg}~(e) to (h)). For non-merging clusters, it is observed that slope is flatter for CRs (i.e. $L_{CR} \propto M^{0.73\pm0.16}$) but steeper in X-rays ($L_X \propto M^{1.40\pm0.04}$). The contrary can be seen in case of merging clusters, where CR luminosity increases very steeply with mass ($L_{CR} \propto M^{1.46\pm0.16}$), while X-ray luminosity scales linearly ($L_X \propto M^{1.02\pm0.06}$). Interestingly, it is observed in the time evolution study in Fig~\ref{merg-phase} that the mass of clusters is not correlated to luminosity during the mass accretion history, especially as far as $L_{CR}$ is concerned.\\

\item Virial ratio as defined in Section~\ref{dyna-stat} has been found to be useful tool for determining the dynamical state of galaxy clusters. From an observational viewpoint, it is sufficient to get the information on velocity dispersion of the constituent galaxies in a cluster, mass of the system, its size, namely the virial radius and second turn over radius to compute its state of virialisation. This quantity exhibits a distinct peak at the time of merger (see Fig.~\ref{vir-ratio}~and~Fig.\ref{merg-phase}) enabling it to be used as a tool to determine the actual dynamical state.\\

\item Simulated merging clusters in our sample do not show virialisation inside the overdensity ratio of r$_{200}$ (see Fig.~\ref{radial-en-ratio}). During the merger phase, the core is dominated by the potential energy, whereas, during the violent relaxation phase, it is mostly dominated by the kinetic energy. A similar finding has been reported in \citet{Shaw2006ApJ}. The apparent discrepancy with the simplified case of the spherical collapse model can be attributed to the complexity of cosmological setup with the presence of filaments and numerous sub-structures and their merger activities.\\

\item To better study the properties of non-virialised objects, we divide them into two categories, HighKE and HighPE (see Section~\ref{dyna-stat} last paragraph for the definitions) the cosmic ray production grows non linearly with mass for HighKE objects ($L_{CR} \propto M^{3.16\pm 0.36}$) compared to HighPE objects ($L_{CR} \propto M^{0.80\pm 0.26}$ and $L_{CR} \propto M^{0.54\pm 0.24}$ with removed outliers). Conversely, the mass scaling of $L_X$ does not show any substantial difference between the two groups. Observed difference of $L_{CR}$ magnitude in HighPE compared to HighKE clusters is a clear indicator of activity for high mass clusters (for mass $> 5\times 10^{14}\odot$). \\

\item CRs produced from the acceleration of injected thermal particles, and the specific case of injection of preCRs described in this study, show a similar nature of production and evolution (both temporal and spatial). PreCR seems important for CR injection by low Mach shocks, during the early stages of mergers and CR content of the inner part ($\sim r_{500}$) of the clusters. PreCR may also be vital for accounting the total CR injection in clusters, which in our special case, show an average increment of half an order in magnitude over the injection from the thermal pool.

\end{itemize}

\section*{Acknowledgements}
This project is funded by DST-SERB, Govt. of India, under the Fast Track scheme for young scientists, Grant No. SR/FTP/PS-118/2011. We are thankful to The Inter-University Centre for Astronomy and Astrophysics (IUCAA) for providing the HPC facility. Computations described in this work were performed using the Enzo code developed by the Laboratory for Computational Astrophysics at the University of California in San Diego (http://lca.ucsd.edu) and data analysis is done with the yt-toolkit (http://yt-project.org/). We would like to thank the anonymous referee for making the most useful comments that helped us to improve the content of the paper significantly. Thanks to Prof. Heysung Kang as well for her valuable suggestions and help regarding the current status of CR acceleration computation.

\bibliographystyle{mn2e}
\bibliography{cosmic_ray_frac_new}

\appendix

\section{Re-acceleration of pre-existing CRs}\label{appen-preCR-study}

As mentioned in Section~\ref{pre-exist}, we have computed  
CR efficiency as a function of Mach number at different state of evolution of our reference cluster $CL_2$, similar to Figure~\ref{mach-cr}. At low Mach numbers (i.e. $\mathcal{M}<3$), preCRs contributions to CR flux is much higher than from only thermal particles. Though the overall nature of the CR flux curves at various stages of mergers is quantitatively similar, with preCRs, clusters produce much higher flux and peaks at lower Mach numbers. While, with preCRs production is most effective between Mach number $\mathcal{M}=2-5.5$ and peak shifts between $\mathcal{M}=2.5-4$. Without preCRs the range is $\mathcal{M}=3-4$ and the effective range is $\mathcal{M}=2.5-5.5$. For strong shocks (beyond $\mathcal{M}>7$), there is no difference in total luminosity for thermal particles and with preCRs.

\begin{figure}
\includegraphics[width=0.52\textwidth]{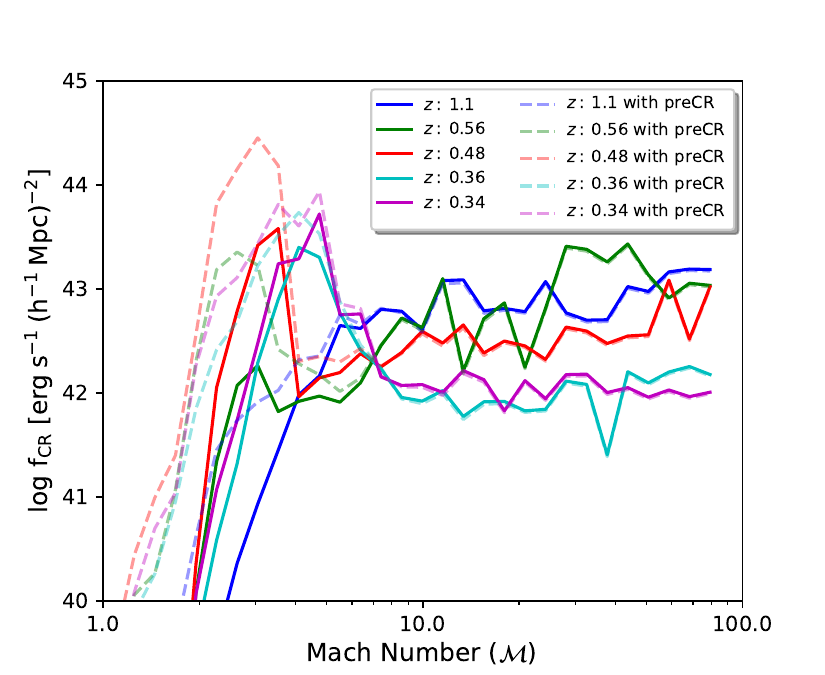}
\caption{Same as Fig.~\ref{mach-cr}, but for preCR (faint dashed line) and thermal particles (deep solid lines).}\label{fig:Lcr-mach}
\end{figure}

\begin{figure}
\includegraphics[width=0.52\textwidth]{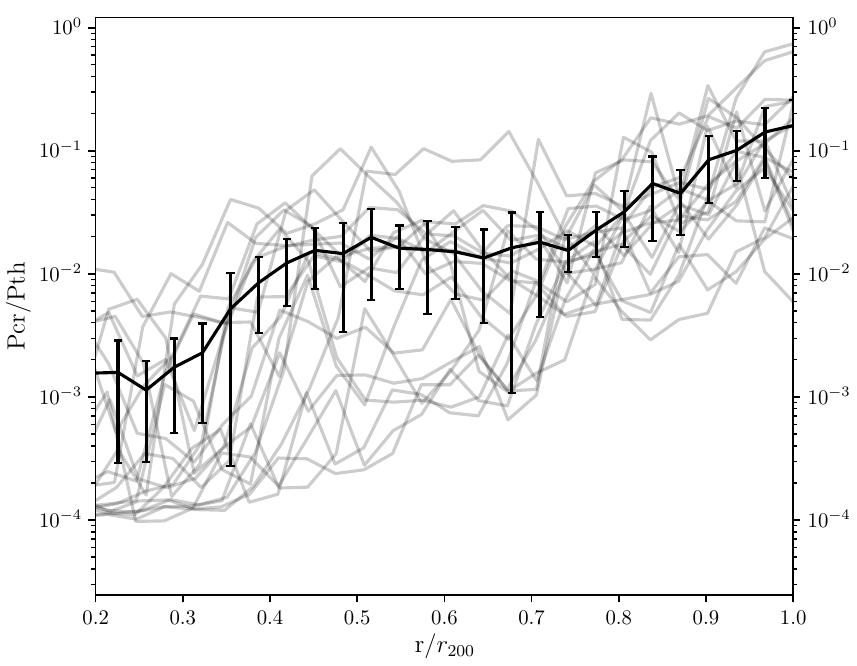}
\caption{Same as Fig.~\ref{fig:pcr_by_pth}, but for preCRs.}\label{fig:PCRbyPTH-pre-ext}
\end{figure}

The radial pressure ratio $P_{CR}/P_{th}$ (as defined in Section~\ref{non-therm} last paragraph) is plotted for the simulated high mass (final mass $> 5\times 10^{14} M_{\odot}$), merging clusters during their evolution after the first merger in Figure~\ref{fig:PCRbyPTH-pre-ext}. It can be noticed in this plot that similar to Figure~\ref{fig:pcr_by_pth}, the ratio goes to as high as 70\% at maximum and beyond radius $r_{500}$ at almost near to $r_{200}$. But, the average profile shows that CR injection pressure is below 1\% in the core and goes to just above 1\% after about 0.3$r_{200}$ till 0.7$r_{200}$. Though assumed initial CR to thermal pressure ratio was 0.05, we see much less radial fraction in the central part of the clusters. This is because of the fact that the CR injection happens only in the effective cells i.e. with Mach number $\mathcal{M}>1.1$ (see section~\ref{cr-compt-model}) and radial profile depends on volume filling as well as strength of shocks, both of which is much less in the core region. Beyond 0.7 (i.e. beyond $r_{500}$), as shock strength as well as volume increase, the ratio rises very fast and it reaches almost 15\% at the outskirts (near to $r_{200}$) of the clusters. When compared the results of preCR with the non-preCR cases as shown in Figure~\ref{fig:pcr_by_pth}, we notice that non-preCR is only effective beyond 0.7 times the virial radius when it crosses 1\% and at virial radius (i.e. $r_{200}$) reaches 10\%. In between 0.3 to 0.7 times the virial radius, preCRs calculation shows the pressure ratio ($P_{CR}/P_{th}$) to be at least a factor of ten larger. 

\begin{figure}
\includegraphics[width=0.52\textwidth]{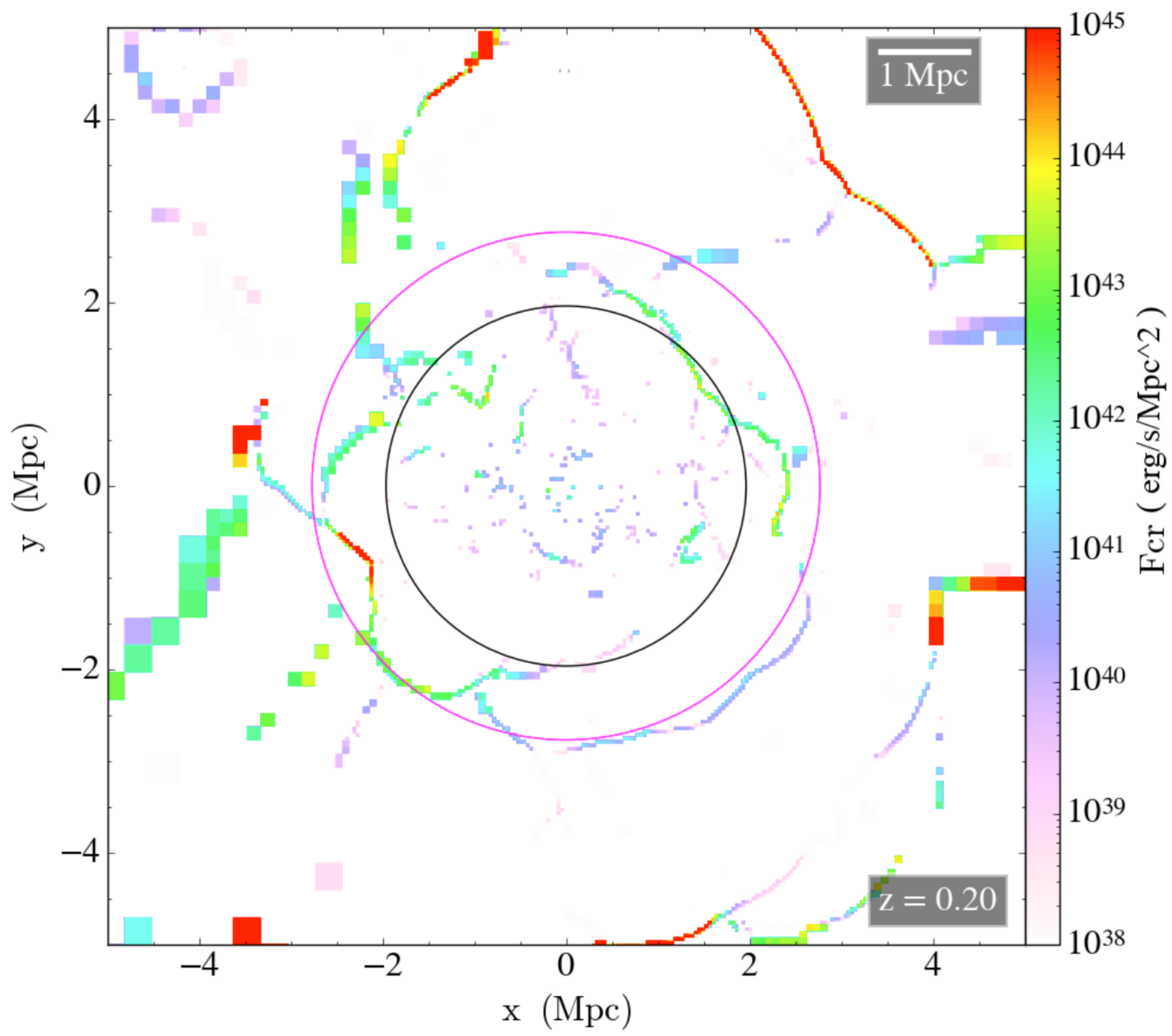}
\caption{Same as Fig.~\ref{fig:mach-CRF}(b), but for preCRs.}\label{fig:Cr-effi-pre-ext}
\end{figure}

Figure~\ref{fig:Cr-effi-pre-ext} is a slice plot of a cluster showing cosmic ray flux when preCRs model is considered. Circles drawn in the image are at $r_{500}$ and at $r_{200}$. A significantly higher flux covering more area can be noticed inside $r_{500}$ compared to non-preCRs model as seen in Figure~\ref{fig:mach-CRF}, though the fluxes and emission area are almost the same beyond $r_{500}$.

\section{Resolution study}\label{appen-res-study}

\begin{figure}
\includegraphics[width=0.52\textwidth]{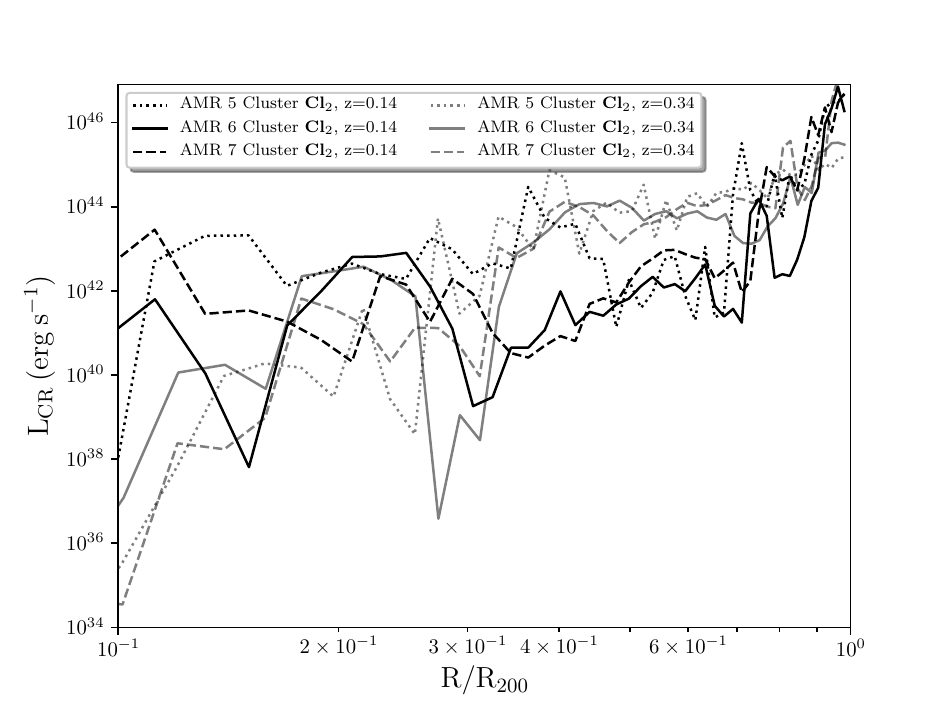}\hspace{-0.4cm}
\hspace{-0.4cm}
\caption{Radial plot of cosmic ray luminosity computed for thermal particles. Radius has been normalised to $r_{200}$ for merging and relaxed states (see Fig~\ref{fig:energy-ev}), of a galaxy cluster ($Cl_2$, mass about 10$^{15} M_{\odot}$) and for three resolutions namely LOWRES, REFRES and HIGHRES (as indicated in the legend respectively). }
\label{radial-res-stud1}
\end{figure}

We show here some resolution studies to test the convergence of our results. The main runs that are used for this study are performed with the cosmological and simulation parameters described in Section~\ref{sample} with 6 levels of total (uni-grid + AMR) refinement leading to a resolution of about 30 kpc. Keeping all other parameters same, we have further simulated some of our objects with different AMR levels to achieve different levels of resolutions. We have produced a lower resolution  (`LOWRES' hereafter, resolution is about 60 kpc) and a higher resolution (`HighRES' hereafter, resolution is about 15 kpc) simulations. Further, to test the convergence of our results, we have performed other two sets of simulations with two different root grids besides the `RefRES'. The one with high resolution root grid i.e. 128$^3$ is named as `RootHIRES' and one with 32$^3$ root grid as `RootLOWRES'. In these test runs, the AMR setup is so chosen that the effective spatial resolution is same as that in reference runs. Finally, we have compared the physical parameters obtained from these simulations.

In Figure~\ref{radial-res-stud1} radial profiles of CR luminosity due to acceleration of thermal particles has been plotted for the clusters $Cl_2$ (merging) and $Cl_5$ (non-merging) in different runs. It can be noticed that the results for the REFRES simulation are almost the same as for the HIGHRES run with minor deviation, though LOWRES data are little away. Small spatial variation that can be noticed among the different resolution plots occurs due to resolution sensitivity of transient phenomena like shocks. It can also be noticed that the merging phases have better convergent results. These results show that our simulated quantities have good convergence with the resolution taken as the reference set of simulations i.e. about 30 kpc with 6 levels of refinement. For further details of resolution study of our data sets, we refer the reader to \citet{Paul_2017MNRAS}.

\begin{figure}
\includegraphics[width=0.52\textwidth]{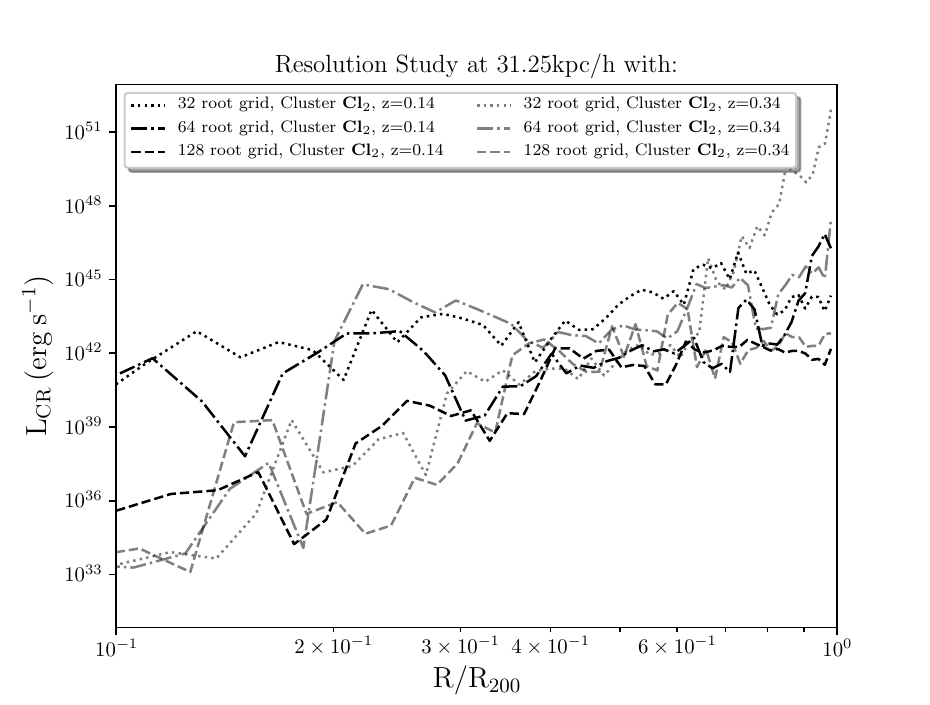}
\caption{Radial profiles of $L_{CR}$, similar to Figure~\ref{radial-res-stud1} for different root grid resolution simulations with the same final resolution of about 30 kpc.}\label{root-res}
\end{figure}

As we mentioned above, other two tests with different root grid resolution namely RootLOWRES and RootHiGHRES have been performed. We have then chosen appropriate merging and non-merging states (as indicated by virial radio $\mathcal{R}_{1}$) and plotted the CR luminosity at those states. The tests with different root grid resolution show a reasonable convergence in CR luminosity (see Figure~\ref{root-res}).

\bsp

\label{lastpage}
\end{document}